\pdfoutput=1
\documentclass[]{aa}  

\usepackage{graphicx}
\usepackage{txfonts}
\usepackage{natbib}
\newcommand{\m}{$\mu m$ }

\DeclareOldFontCommand{\rm}{ }{\mathrm}
\everymath{\rm}

\begin{document}

   \title{Kinematic and Thermal Structure at the onset of high-mass star formation\thanks{FITS files of Figs. \ref{IRDC01070_overview} to \ref{ISOSS23053_overview} are available in electronic form at the CDS via anonymous ftp to cdsarc.u-strasbg.fr (130.79.128.5)}}

   \subtitle{}

   \author{S. Bihr\inst{1}
          \and
          H. Beuther\inst{1}
          \and
          H. Linz\inst{1}
          \and
          S. E. Ragan\inst{1}
          \and
          M. Hennemann\inst{2}
          \and
          J. Tackenberg\inst{1}
          \and
          R. J. Smith\inst{3}
          \and
          O. Krause\inst{1}          
          \and
          Th. Henning\inst{1}
          }

   \institute{Max Planck Institute for Astronomy, K\"onigstuhl 17, 69117 Heidelberg, Germany\\
              \email{name@mpia.de}
    \and
    Laboratoire AIM, CEA/DSM-CNRS-Université Paris Diderot, IRFU/Service d’Astrophysique, CEA Saclay, 91191 Gif-sur-Yvette, France
    \and
    Institut f\"ur Theoretische Astrophysik, Zentrum f\"ur Astronomie, Universit\"at Heidelberg, Albert-Ueberle-Str. 2, D-69120 Heidelberg, Germany
             }
   \date{Received February 08, 2013; accepted February 20, 2015}
  \abstract
   {Even though high-mass stars are crucial for understanding a diversity
of processes within our galaxy and beyond, their formation and initial conditions
are still poorly constrained.}
   {We want to understand the kinematic and thermal properties of young massive
gas clumps prior to and at the earliest evolutionary stages of high-mass
star formation. Do we find signatures of gravitational collapse? Do we find
temperature gradients in the vicinity or absence of infrared emission sources?
Do we find coherent velocity structures toward the center of the dense and
cold gas clumps?}
   {To determine kinematics and gas temperatures, we used ammonia, because
it is known to be a good tracer and thermometer of dense gas. We observed
the NH$_3$ (1,1) and (2,2) lines within seven very young high-mass star-forming
regions comprised of infrared dark clouds (IRDCs), along with ISO-selected
far-infrared emission sources (ISOSS) with the Karl G. Jansky Very Large
Array (VLA) and the Effelsberg 100m telescope. }
   {The molecular line data allows us to study velocity structures, linewidths,
and gas temperatures at high spatial resolution of 3-5$''$, corresponding
to $\sim 0.05~$pc at a typical source distance of 2.5~kpc. We find on average cold
gas clumps with temperatures in the range between 10~K and 30~K. The observations
do not reveal a clear correlation between infrared emission peaks and ammonia
temperature peaks. Several infrared emission sources show ammonia temperature
peaks up to 30~K, whereas other infrared emission sources show no enhanced
kinetic gas temperature in their surrounding. We report an upper limit for
the linewidth of $\sim 1.3~km\, s^{-1}$, at the spectral resolution limit
of our VLA observation. This indicates a relatively low level of turbulence
on the scale of the observations. Velocity gradients are present in almost
all regions with typical velocity differences of $1$ to  $2~km\, s^{-1}$
and gradients of $5$ to $10~km\, s^{-1}\, pc^{-1}$. These velocity gradients
are smooth in most cases, but there is one exceptional source (ISOSS23053),
for which we find several velocity components with a steep velocity gradient
toward the clump centers that is larger than $30~km\,s^{-1}\,pc^{-1}$. This
steep velocity gradient is consistent with recent models of cloud collapse.
Furthermore, we report a spatial correlation of ammonia and cold dust, but
we also find decreasing ammonia emission close to infrared emission sources.}
   {}

   \keywords{Stars: formation - Stars: massive - ISM: clouds - ISM: kinematics
and dynamics - Techniques: interferometric}

   \maketitle

\section{Introduction}

\begin{table*}
\caption{Sample details}             
\label{sample_details}      
\centering                
\begin{tabular}{c c c c c c c c c c}        
\hline\hline                
Source & $\alpha$ (J2000) & $\delta$ (J2000) & v$_{lsr}$ & D\tablefootmark{a}&
rms & Beam size  & $\langle$T$_K\rangle$ & F$_{SCUBA}$ & M \\
name &  $(h:m:s)$ & $(^{\circ} : ' : '')$  & ($km\,s^{-1}$) &  (kpc)& (mJy)
& ($''$ x $''$)  & (K) & (Jy) & (M$_\odot$)\\   

\hline                    
 
IRDC010.70 & 18:09:45.9 & -19:42:12  & +28.8 & 3.5 $\pm\:0.5$& 3.0 & 5.7
x 3.3 & \tablefootmark{b} & 6.6 $\pm\:1.3$ & 820\tablefootmark{c}\\

IRDC079.31 & 20:31:58.0 & +40:18:20  & +1.3 &  1.6 $^{+\:2.0}_{-\:1.6}$&
3.2 & 5.1 x 2.9 & 12.8  & 5.9 $\pm\:1.2$ & 200\\

ISOSS18364 & 18:36:36.0 & -02:21:45  & +35.0 &  2.4 $\pm\:0.4$& 2.9 & 4.6
x 3.1 & 12.7 & 1.4 $\pm\:0.3$  & 110\\

ISOSS20153 & 20:15:21.7 & +34:53:50  & +2.5 &  1.2 $\pm\:1.1$& 2.9 & 4.7
x 2.9 & 17.0 & 2.1 $\pm\:0.4$& 25 \\

ISOSS22478 & 22:47:49.6 & +63:56:45  & -39.7 &  3.2 $\pm\:0.6$& 3.4 & 5.0
x 3.0 & 13.2\tablefootmark{d} & 1.1 $\pm\:0.2$& 140\\

ISOSS23053 & 23:05:22.5 & +59:53:50  & -51.7 &  4.3 $\pm\:0.6$& 3.4 & 5.1
x 3.1  & 18.3 & 4.4 $\pm\:0.9$ & 610\\
\hline                  
\end{tabular}

\tablefoot{Source name, coordinates, velocities, distances, rms of line free
channels, synthesized beam size, average temperatures from the Effelsberg
100~m telescope observations, SCUBA-fluxes at 850~$\mu m,$ and clump masses
of our sample.\\
\tablefoottext{a}{Distances are taken from \citet{Ragan2012b}}\\
\tablefoottext{b}{No Effelsberg data available.}\\
\tablefoottext{c}{Used T = 15K to calculate mass.}\\
\tablefoottext{d}{Temperature is an upper limit, because of non-detection
of NH$_3$(2,2) line.}\\}
\end{table*}

Even though the understanding of high-mass star formation has made tremendous
progress over the past decade \citep{Beuther2007,Zinnecker2007,Klessen2011,Tan2014},
the initial conditions are still poorly constrained. It is known, that stars
predominantly form in clusters \citep{Lada2003}, which is especially true
for high-mass stars \citep[e.g.,][]{Wit2005, Gvaramadze2012}. On observational
grounds, the first evolutionary stage of such clusters spawning future high-mass
stars might in general be termed pre-protocluster cores \citep{Evans2002}.
Their appearance and morphology can be different depending on the environment.
Objects of this sort have been found in several (sub-)millimeter surveys
\citep[e.g.,][]{Klein2005, Beuther2007b}. Also the ISO satellite mission
has resulted in a list of such objects revealed by far-infrared observations
at 170~$\rm{\mu}$m within the ISO Serendipity Survey (ISOSS, \citet{Bogun1996}).
For these clumps, cold dust and gas temperatures have been established in
follow-up investigations \citep{Krause2003, Krause2004}. The most prominent
variety of young massive clumps are the infrared dark clouds (IRDCs). They
were discovered as dark silhouettes against the galactic background at 8
and 15~$\rm{\mu}$m with the Midcourse Space Experiment \citep[MSX,][]{Egan1998}
and ISO \citep{Perault1996}. ISO, MSX, the Spitzer Space telescope, and further
(sub-)millimeter observations have helped for studying these objects in detail
\citep[e.g.,][]{Simon2006,Hennemann2008,Vasyunina2009,Ragan2009,Peretto2009}.
Because IRDCs can only be seen in absorption against a strong infrared background,
their location is mainly within the disk toward the inner quadrants of our
Milky Way, whereas ISOSS sources are more widely distributed. In fact, the
Milky Way midplane had to be avoided for ISOSS because of saturation. IRDCs
can have masses up to several thousand solar masses, and the more massive
ones are explicitly thought to be progenitors of star clusters \citep[e.g.,][]{Wyrowski2008,Rathborne2010,Henning2010,Ragan2012b}.
On average the ISOSS sources have masses less than 500~M$_\odot$ \citep{Hennemann2008}.
While their masses often are somewhat lower than for large high-contrast
IRDCs, the general characteristics such as low temperature and low turbulent
linewidth are similar to IRDCs. Therefore, ISOSS sources are also considered
to be early evolutionary stages of intermediate- to high-mass star formation
in more isolated regions. Consequently, the better characterization of such
clumps (ISOSS and IRDCs) is an important step toward understanding the initial
conditions of high-mass star formation. \\
With the Herschel observatory \citep{Pilbratt2010}, it has been possible
to observe the dust content of star-forming regions in unprecedented detail
at far-infrared wavelengths. To study the onset of star formation, we conducted
the Herschel Guaranteed Time Key Project EPoS - \textit{the Earliest Phases
of Star formation}. The goal of this project was to map 45 high-mass star-forming
regions \citep{Ragan2012b}, as well as 15 low-mass globules \citep{Launhardt2013}
in the far-infrared continuum emission. The sample of 45 high-mass regions
includes regions from the subsamples mentioned above: ISOSS sources and IRDCs.
An overview of the high-mass regions of the EPoS project is presented in
\citet{Ragan2012b}. They find almost 500 point sources within 45 regions
and characterize their structure and temperature. More than half of them
have 24~$\rm{\mu}$m counterparts. They conclude that the presence of a 24~$\rm{\mu}$m source implies an embedded protostar.\\
The downside of continuum observations, such as the observations with the
Herschel's bolometer cameras within PACS and SPIRE, is the lack of kinematic
information. As the formation process of stars is thought to be highly dynamic
\citep[e.g.,][]{Bonnell2004,Klessen2005,Csengeri2011b}, such information
is crucial for constraining star formation theories. As a result, follow-up
observations of molecular lines are mandatory for investigating the kinematics
of star-forming regions.\\
Early interferometric measurements toward IRDCs in the (sub-)millimeter range
often concentrated on embedded infrared-bright protostars \citep[e.g.,][]{Redman2003,Pillai2006b}
or in continuum studies that investigate the clump and core fragmentation
\citep[e.g.,][]{Rathborne2007}. Fewer (sub-)millimeter interferometric studies
report on spectroscopic observations. But the interferometric detection of
molecular lines (sometimes even CO) away from embedded infrared sources has
been difficult \citep[e.g.,][]{Rathborne2008,Zhang2009}. More recent interferometric
studies, with emphasis on investigating the fragmentation of IRDCs, often
use N$_2$H$^+$ as a tracer \citep[e.g.,][]{Beuther2009,Kauffmann2013,Beuther2013}.
However, it is difficult to measure the temperature distribution of IRDCs
at high angular resultion, since many of the usual temperature tracers in
the millimeter, such as CH$_3$CN or H$_2$CO, are problematic when it comes
to the low-temperature regime of IRDCs (too low excitation and/or density,
freeze-out, etc.). Therefore, measurements of the NH$_3$ inversion lines
are ideal for temperature maps of IRDC-like environments \citep{Ho_1983,
Ragan2011, Pillai2006}.\\
For this project, we observed ammonia, not just because it is a good thermometer,
but it is also known to be a good tracer of dense gas and can be used to
study kinematics \citep{Pillai2006,Ragan2012}. As the density of high-mass
star-forming regions is in the range $10^4 - 10^5~cm^{-3}$ \citep{Beuther2007},
ammonia traces these clumps well. Several ammonia surveys have been conducted
recently, using single-dish telescopes with a spatial resolution of $>40''$
\citep{Pillai2006, Dunham2011, Wienen2012, Purcell2012, Chira2013} and interferometers
at high spatial resolution of $< 5''$ \citep{Wang2008, Ragan2011, Devine2011,
Ragan2012, Brogan2011, Zhang2011, Lu2014, Sanchez2013}. The observed objects
vary in terms of mass, distance, evolutionary stages, and infrared emission.
The results and conclusions are therefore diverse. For example, \citet{Wang2008}
reports decreasing rotational temperature toward the center of the IRDC G28.34+0.06,
and the sample of \citet{Ragan2012} shows active collapse and fragmentation.\\
Our goal is to resolve the kinematic and temperature structure within a sample
of very young high-mass star-forming regions at high spatial resolution of
$\sim 3-5''$, which results in a linear resolution of $\sim 10000~AU \approx
0.05~pc$ at typical distances of 2.5~kpc. While our target regions are part
of IRDC and ISOSS sources, here we concentrate on the substructures within
them. To distinguish between clouds, clumps, and cores, we follow the nomenclature
of \citet{Bergin2007}. According to this, cores, clumps, and clouds have
sizes of $<0.2~pc$, $0.3-3~pc,$ and $2-15~pc$, respectively. Similar classifications
can be found in the literature \citep[e.g.,][]{Williams2000, Beuther2007,
Klessen2011}. Clumps are thought to be progenitors of star clusters, whereas
cores form single or binary stars. Since our objects have sizes between $\sim
0.3~pc$ and $1~pc$, we classify them as clumps. Within these clumps, we are
able to resolve individual cores, which may form single or binary stars.
Important questions that we address in this paper are: What is the temperature
structure at the onset of high-mass star formation? Do we see temperature
drops toward the highest density peaks similar to low-mass star formation?
How dominant is turbulence on the given scale? What are the dynamical signatures
of cloud or clump collapse? Can we identify infall signatures on the small
scales of individual cores?\\ 
In Section \ref{observation_and_data_reduction}, we explain our observations
with the Karl G. Jansky Very Large Array (VLA) and the Effelsberg 100m telescope.
Our observational results are explained in Section \ref{results}, and we
discuss them in Section \ref{discussion}. Section \ref{conclusions} provides
a summary and our conclusions.

\begin{table*}
\caption{Physical properties of cores}             
\label{core_details}      
\centering                
\begin{tabular}{c c c c c c c c c c}        
\hline\hline                 

Source & ID & $\alpha$ (J2000) & $\delta$ (J2000) & S$_{24}$ & S$_{70}$ &
T$_{dust}$ & L & M \\
name & number & (h:m:s) & $(^{\circ} : ' : '')$ & (Jy) & (Jy) & K & L$_\odot$
& M$_\odot$\\   
\hline
IRDC079.31 & 12 & 20:31:58.3 & 40:18:38 & 0.23 & 1.68 & 18 & 25 & 7 \\
IRDC010.70 & 10 & 18:09:45.7 & -19:42:08 & 0.04 & 2.11 & 20 & 98 & 17 \\
ISOSS18364 & 2 & 18:36:36.2 & -2:21:49 & 0.10 & 14.96 & 24 & 173 & 11 \\
ISOSS20153 & 1 & 20:15:21.3 & 34:53:45 & 3.32\tablefootmark{a} & 20.83 &
24 & 65 & 4 \\
ISOSS22478 & 1 & 22:47:46.6 & 63:56:49 & -\tablefootmark{b} & 0.16 & 18 &
10 & 3 \\
ISOSS22478 & 2 & 22:47:46.8 & 63:56:32 & 0.10 & 0.28 & 21 & 7 & 0.8 \\
ISOSS22478 & 3 & 22:47:48.2 & 63:56:44 & -\tablefootmark{b} & 0.31 & 21 &
9 & 1 \\
ISOSS22478 & 4 & 22:47:50.1 & 63:56:45 & 0.11 & 0.22 & 20 & 7 & 1 \\
ISOSS22478 & 5 & 22:47:50.6 & 63:56:57 & -\tablefootmark{b} & 0.27 & 21 &
7 & 0.9 \\
ISOSS23053 & 2 & 23:05:21.7 & 59:53:42 & 0.10 & 7.63 & 21 & 441 & 56 \\ 
ISOSS23053 & 3 & 23:05:23.7 & 59:53:53 & 0.51 & 18.16 & 22 & 869 & 84 \\

\hline                  
\end{tabular}

\tablefoot{Core properties investigated by \citet{Ragan2012b}. The ID number
describes the core number of each clump specified in \citet{Ragan2012b}.
S$_{24}$ is Spitzer MIPS 24~\m and S$_{70}$ is Herschel PACS 70. Temperature,
mass, and luminosity are calculated from a best-fit spectral energy distribution
of the Herschel infrared data using the PACS 70, 100, 160~$\mu$m data.
Details can be found in \citet{Ragan2012b}.\\
\tablefoottext{a}{IRAS 25~$\rm{\mu}$m flux.}\\
\tablefoottext{b}{Not detected.}\\
}
\end{table*}

\section{Observations and data reduction}
\label{observation_and_data_reduction}
We observed the NH$_3$ (1,1) and (2,2) inversion lines of a subset of the
EPoS high-mass sample \citep{Ragan2012b} with the VLA and Effelsberg 100m
telescopes. The rest frequency of the NH$_3$ (1,1) and (2,2) inversion lines
are 23.694495~GHz and 23.722633 GHz \citep{Ho_1983}, respectively. 

\subsection{Sample}

The observed sample consists of two IRDCs and four ISOSS sources. Their kinematic
distances range between $\sim 1-4~kpc$ and are taken from \citet{Ragan2012b}.
They calculated the distance using the method reported in \citet{Reid2009}.
The estimated mass has a wide spread between $\sim 30$ and $800~M_\odot$
(see Section\ref{cloud_masses}). The sample details can be found in Table
\ref{sample_details} and include v$_{lsr}$, the estimated rms, the synthesized
beam size, the distance, the average temperature, the measured SCUBA flux
at 850~$\mu m,$ and the mass estimated from the SCUBA data. We used the SCUBA
data to determine the mass (see section \ref{cloud_masses}). \citet{Ragan2012b}
investigated the infrared point sources within the EPoS sample. Our sample
includes some of these cores, and their details are given in Table \ref{core_details}.

\subsection{Very Large Array observations}

The VLA observations were conducted in July and August 2010 during the VLA
upgrade as shared risk observations. We used the array in D configuration,
achieving a synthesized beam of $3-5''$. As a flux and bandpass calibrator,
we observed J1331+3030 and J1733-1304, respectively. Phase calibrations were
done with regular observations of J2025+3343, J2322+5057, J1743-0350, and
J1851+0035 every $\sim 5$~min and pointing every $\sim 30$~min.\\
Because the WIDAR correlator was used to observe the NH$_3$ (1,1) and (2,2)
lines simultaneously, we could avoid most relative calibration errors. A
bandwidth of 4~MHz and 64 channels resulted in a channel width of 62.5~kHz.
This corresponds to a spectral resolution of $\sim 0.8~km\,s^{-1}$. The bandwidth
of 4~MHz gives us a velocity range of $\sim 50~km\,s^{-1}$. The low sensitivity
of the edge channels meant that we are not able to use $\sim$5$~\%-$10$~\%$
of the edge channels. We therefore can use about 55 channels for our analysis,
which gives us a velocity range of $\sim 44~km\,s^{-1}$. This allows us
to observe both the inner satellite and one outer satellite of the NH$_3$(1,1)
hyperfine structure (see Figs.~\ref{appendix_spectra_IRDC01070}-\ref{appendix_spectra_ISOSS23053}).
To achieve a $1\sigma$ noise level of $\sim3~mJy\,beam^{-1}$, we integrated
30~min per source.\\
The VLA data were calibrated, reduced, and cleaned with the CASA 3.3 package
(Common Astronomy Software Applications), provided by NRAO.

\subsection{Effelsberg 100m Telescope observations}

To overcome missing-flux problems and to reconstruct large scale structures,
we observed our sample with the Effelsberg 100m Telescope in January and
February 2012 (except IRDC010.70 since it is not ideal to observe with the
Effelsberg 100m telescope owing to low declination). We observed raster maps
in frequency-switching mode, covering the whole VLA primary beam of $\sim
120''$. The primary beam of Effelsberg at $1.3~cm$ is $40''$, and with an
approximate Nyquist-sampling of $\sim 20'',$ this requires map sizes of $7\times
7$ positions. For the smaller objects in our sample (ISOSS18364, ISOSS20153),
we observed $5\times 5$ positions. We used the new fast Fourier transform
spectrometer (xFFTS) with a bandpass of 100 MHz and 32768 channels, centered
at 23708.5646~MHz. This setup allowed us to observe the NH$_3$ (1,1) and
(2,2) lines (including the full hyperfine structure) simultaneously, hence
to eliminate most relative calibration errors. The spectral resolution was
$\sim 0.04~km\,s^{-1}$. As a flux calibrator, we used NGC7027 \citep{Zijlstra2008}.

With an integration time of 4~min per raster position, we achieved a  $1\sigma$
noise level of $\sim20-70~mJy\,beam^{-1}$, dependent on the declination
of the source, multiple coverage, and weather conditions. The data calibration
and imaging was done with GREG and CLASS of the GILDAS package by IRAM.

\subsection{Combination of single-dish and interferometry data}

We used the \textit{\emph{feather}} task in CASA to combine the VLA and Effelsberg
100m Telescope data. The Effelsberg 100m Telescope data had to be smoothed
to the spectral resolution of the VLA ($\sim 0.8~km\,s^{-1}$). The combined
data includes the large scale structure ($\sim$40\arcsec) observed with the
Effelsberg 100m telescope, as well as the small scale structure ($\sim$3-5\arcsec)
observed with the VLA. The missing flux in the VLA data depends on the source
structure, sidelobes, and integration area and ranges between $\sim$40-95$\%$.

\begin{figure*}
    \centering
       \includegraphics[width=17cm]{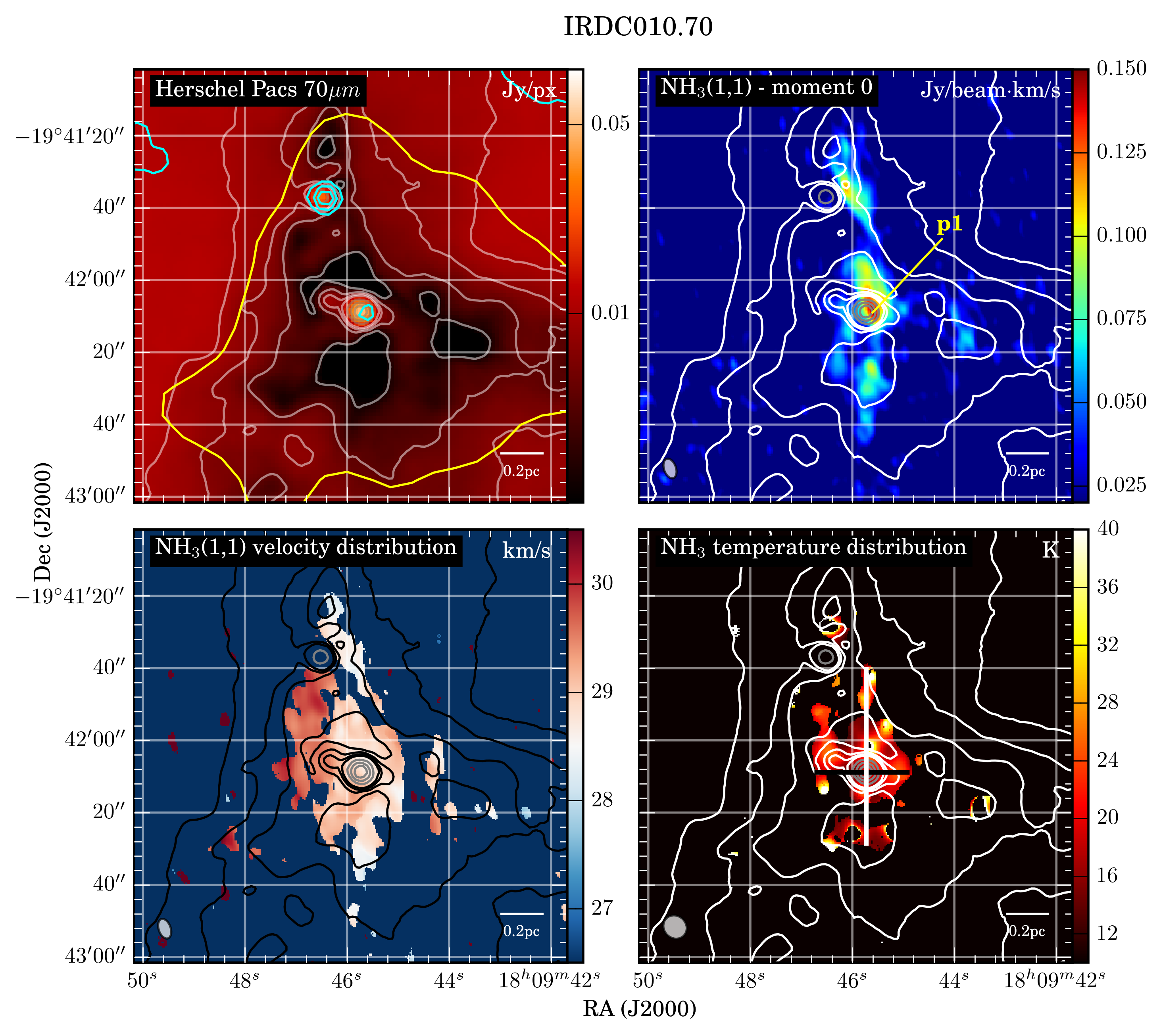}
       \caption{IRDC010.70 - Overview. The upper left panel shows Herschel
PACS 70~$\mu m$ emission and cyan contours present Spitzer MIPS 24~$\mu m$
emission at levels of 90, 110, 130, and 150~MJy/sr. The yellow contour shows
the 3-sigma limit of the SCUBA 850~$\mu m$ emission and therefore the area
where we measured the clump mass. The following panels present the VLA data.
The upper right panel shows the integrated emission of the main NH$_3$(1,1)
line in $Jy\, beam^{-1}\, km\, s^{-1}$ from $27.1$ to $30.3~km\, s^{-1}$.
The lower left panel presents the peak velocity distribution in $km\, s^{-1}$.
The lower right panel shows the kinetic temperature in K. We extracted the
temperature distribution along the black and white lines shown in the lower
right panel and plotted the values as a function of distance in Fig. \ref{fig_temp_cut}.
In each panel contours present Herschel PACS 70~$\mu m$ emission at levels
of 2.5, 5, 7.5, and 10~mJy/px for white/black contours and 20, 30, 40, and
50~mJy/px for gray contours. The NH$_3$(1,1) emission peak is marked in the
upper right panel, and the corresponding spectra are given in Fig.\ref{appendix_spectra_IRDC01070}.
The synthesized beam and scale are shown in each panel.}
        \label{IRDC01070_overview}
\end{figure*}

\begin{figure*}
    \centering
       \includegraphics[width=17cm]{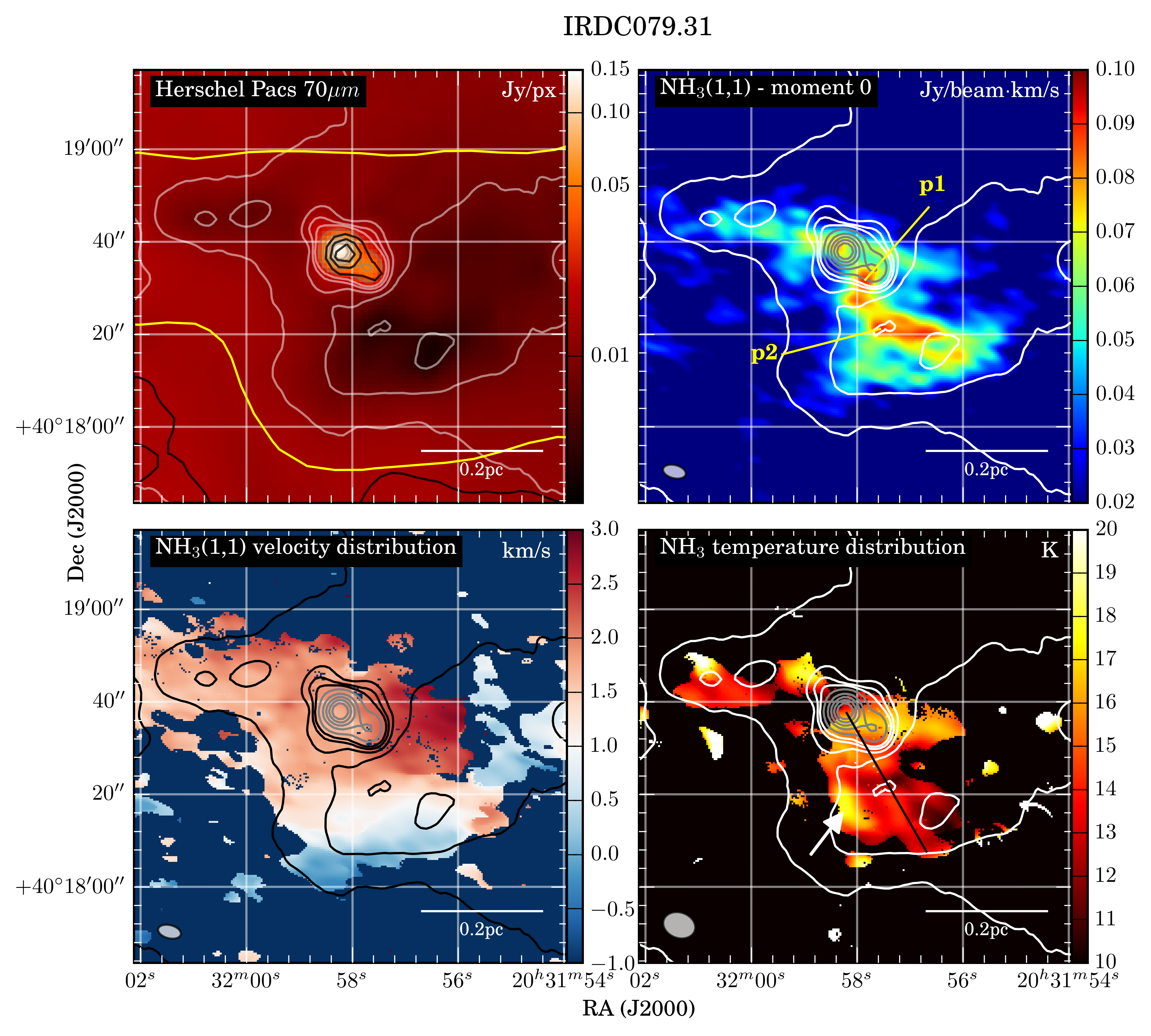}
       \caption{IRDC079.31 - Overview. The upper left panel shows Herschel
PACS 70~$\mu m$ emission and black contours present Spitzer MIPS~24 $\mu m$
emission at levels of 80, 115, and 150~MJy/sr. The yellow contour shows the
3-sigma limit of the SCUBA 850~$\mu m$ emission and therefore the area where
we measured the clump mass. The following panels present the VLA and Effelsberg
100m telescope data combined. The upper right panel shows the integrated
emission of the main NH$_3$(1,1) line in $Jy\, beam^{-1}\, km\, s^{-1}$ from
$-0.3$ to $2.8~km\, s^{-1}$. The lower left panel presents the peak velocity
distribution in $km\, s^{-1}$. The lower right panel shows the kinetic temperature
in K. We extracted the temperature distribution along the black line shown
in the lower right panel and plotted the values as a function of distance
in Fig. \ref{fig_temp_cut}. In each panel contours present Herschel PACS
70~$\mu m$ emission at levels of 4, 8, 12, 16, and 20~mJy/px for white/black
contours and 30, 50, 70, 90, and 110~mJy/px for gray contours. The NH$_3$(1,1)
emission peaks are marked in the upper right panel, and the corresponding
spectra are given in Fig.\ref{appendix_spectra_IRDC079.31}. The synthesized
beam and scale are shown in each panel.}
        \label{IRDC07931_overview}
\end{figure*}

\begin{figure*}
 \centering
 \includegraphics[width=17cm]{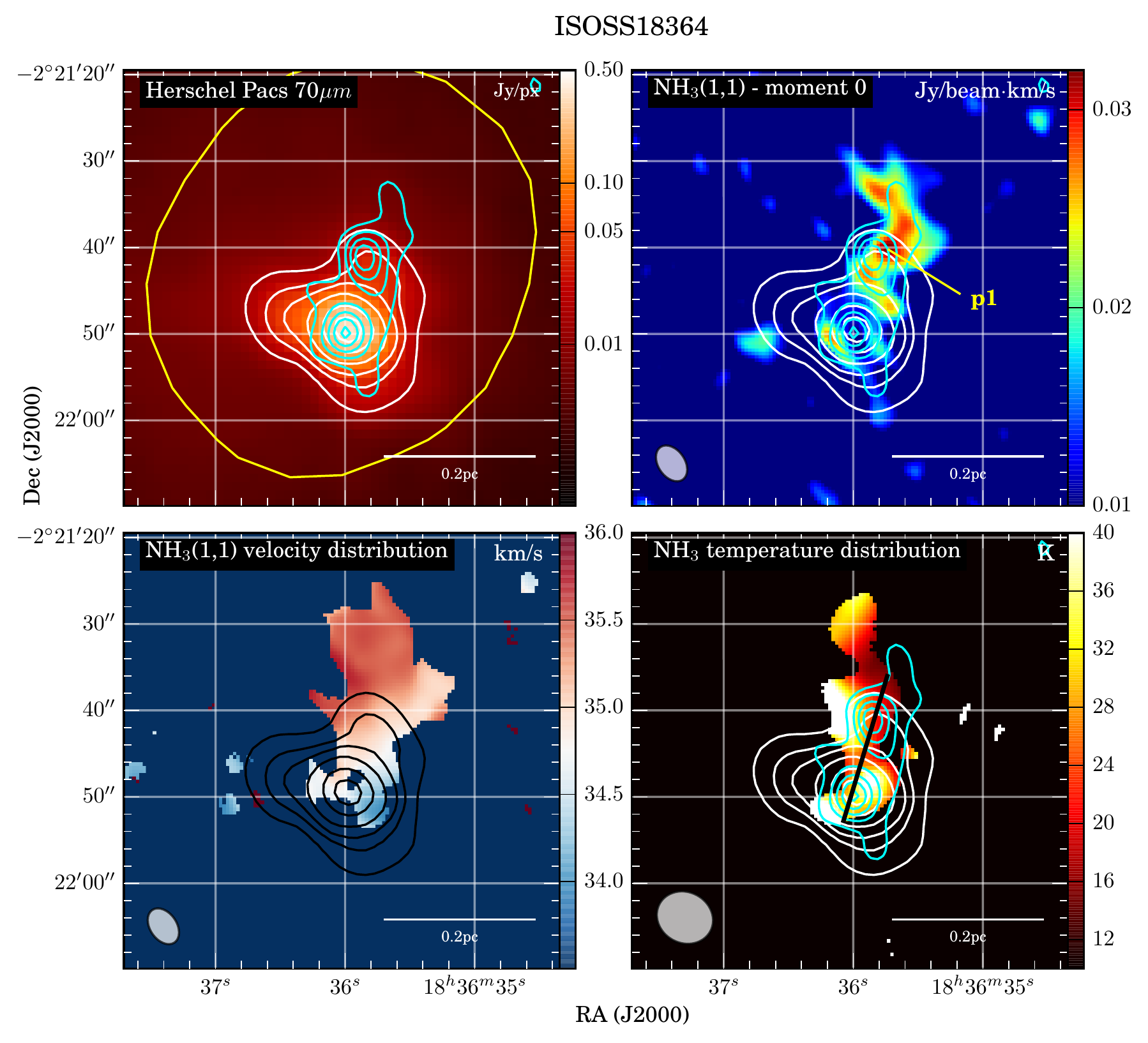}
 \caption{ISOSS18364 - Overview. The upper left panel shows Herschel PACS
70~$\mu m$ emission. The yellow contour shows the 3-sigma limit of the SCUBA
850~$\mu m$ emission and therefore the area where we measured the clump mass.
The following panels present the VLA and Effelsberg 100m telescope data combined.
The upper right panel shows the integrated emission of the main NH$_3$(1,1)
line in $Jy\, beam^{-1}\, km\, s^{-1}$ from $34.2$ to $36.6~km\, s^{-1}$.
The lower left panel presents the peak velocity distribution in $km\, s^{-1}$.
The lower right panel shows the kinetic temperature in K. We extracted the
temperature distribution along the black line shown in the lower right panel
and plotted the values as a function of distance in Fig. \ref{fig_temp_cut}.
In each panel white/black contours present Herschel PACS 70~$\mu m$ emission
at levels of 15, 30, 50, 100, 200, and 300~mJy/px and cyan contours present
PdBI continuum 3.4~mm at levels of 0.4, 0.8, 1.2, 1.6, and 2~mJy/beam with
a synthesized beam of $4.6'' \times 3.2''$ (\cite{Hennemann2009}). The NH$_3$(1,1)
emission peak is marked in the upper right panel, and the corresponding spectra
are given in Fig.\ref{appendix_spectra_ISOSS18364}. The synthesized beam
and scale are shown in each panel.}
 \label{ISOSS18364_overview}
\end{figure*}

\begin{figure*}
 \centering
 \includegraphics[width=17cm]{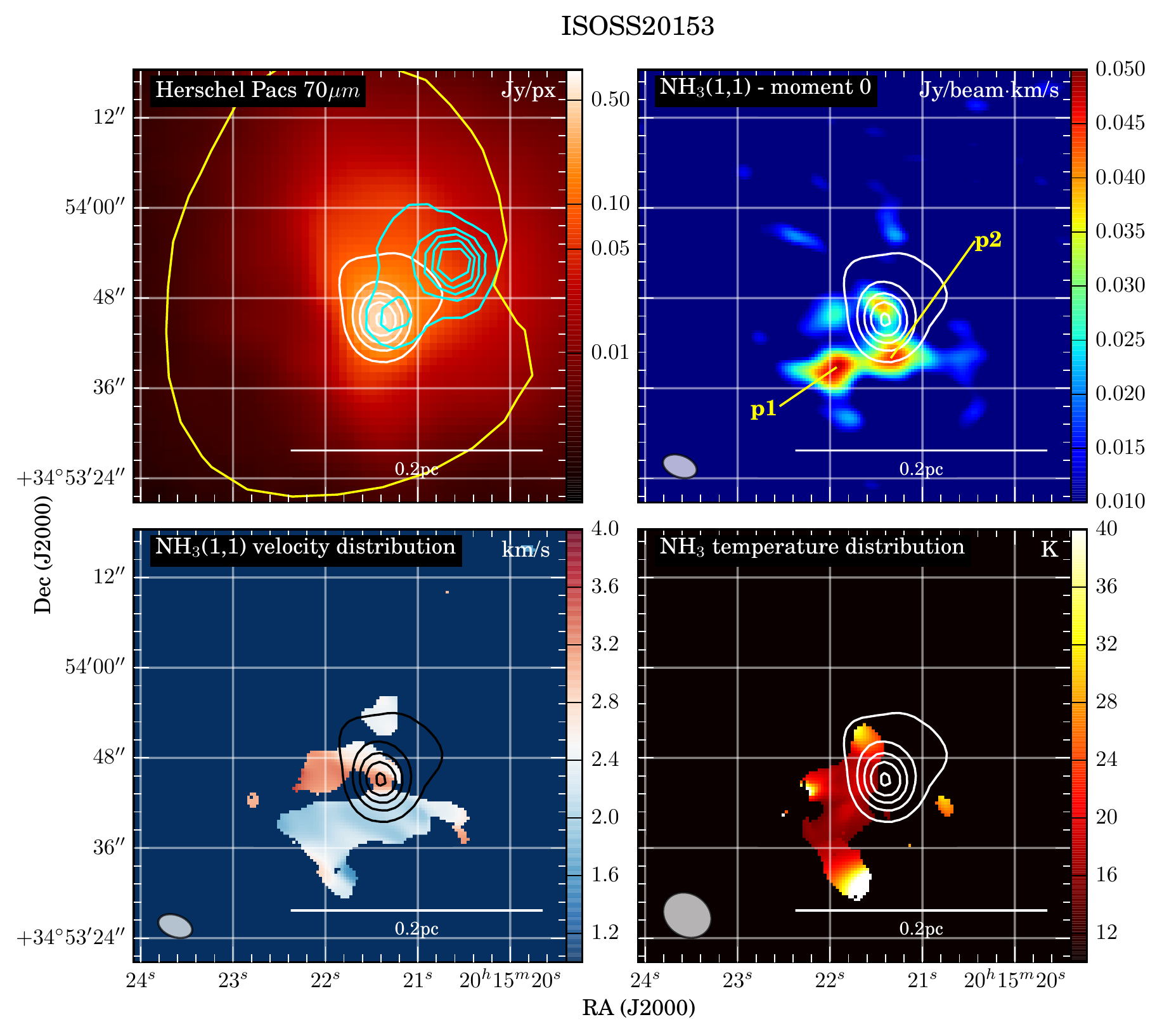}
 \caption{ISOSS20153 - Overview. The upper left panel shows Herschel PACS
70~$\mu m$ emission and cyan contours present Spitzer MIPS 24~$\mu m$ emission
at levels of 150, 250, 350, 450, and 550~MJy/sr. The yellow contour shows
the 3-sigma limit of the SCUBA 850~$\mu m$ emission and therefore the area
where we measured the clump mass. The following panels present the VLA and
Effelsberg 100m telescope data combined. The upper right panel shows the
integrated emission of the main NH$_3$(1,1) line in $Jy\, beam^{-1}\, km\,
s^{-1}$ from $0.8$ to $4.0~km\, s^{-1}$. The lower left panel presents the
peak velocity distribution in $km\, s^{-1}$. The lower right panel shows
the kinetic temperature in K. In each panel white/black contours present
Herschel PACS 70~$\mu m$ emission at levels of 0.1, 0.2, 0.3, 0.4, and 0.5~Jy/pixel. The NH$_3$(1,1) emission peaks are marked in the upper right panel,
and the corresponding spectra are given in Fig.\ref{appendix_spectra_ISOSS20153}.
The synthesized beam and scale are shown in each panel.}
 \label{ISOSS20153_overview}
\end{figure*}

\begin{figure*}
 \centering
 \includegraphics[width=17cm]{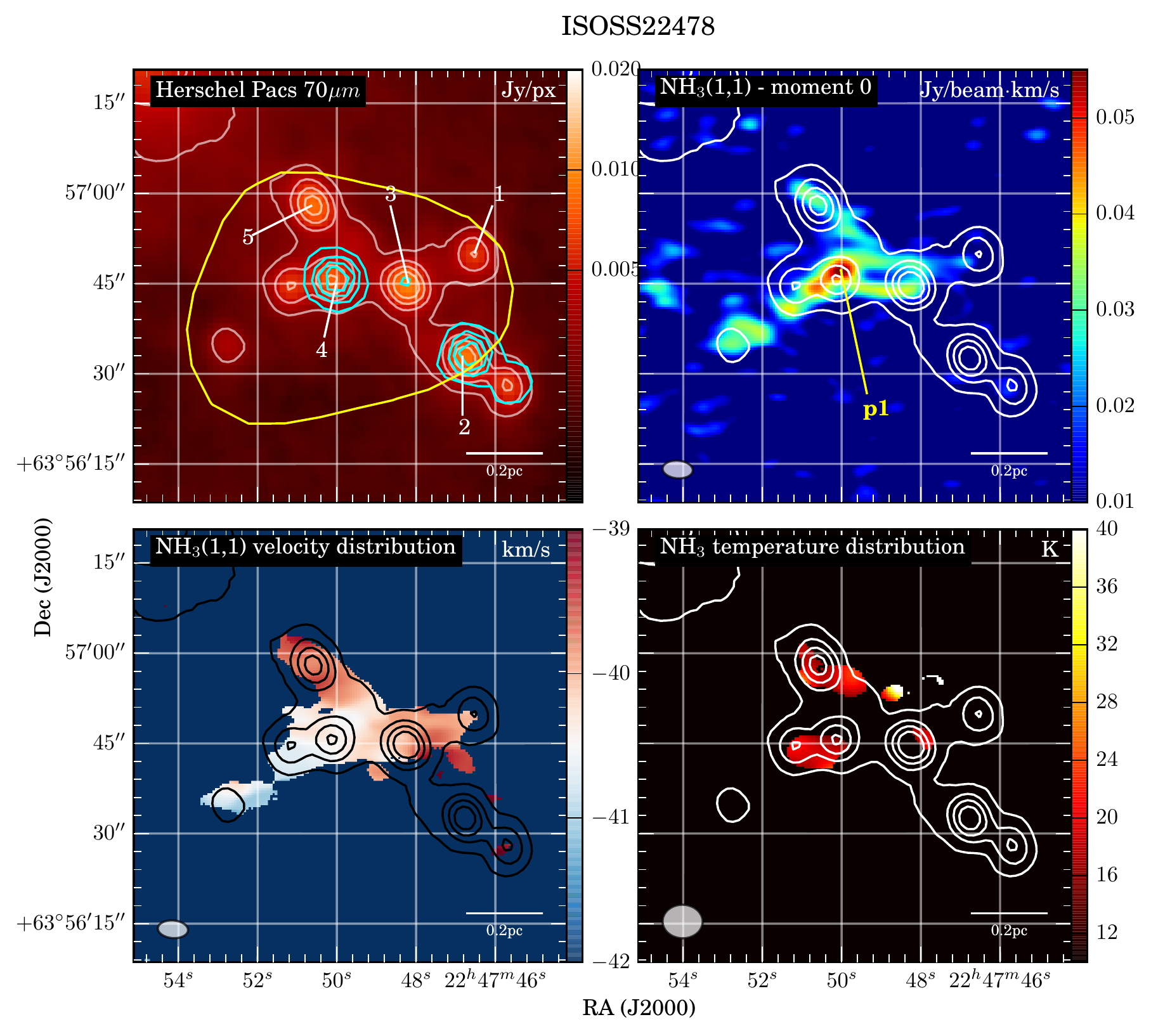}
 \caption{ISOSS22478 - Overview. The upper left panel shows Herschel PACS
70~$\mu m$ emission, and cyan contours present Spitzer MIPS 24~$\mu m$ emission
at levels of 15, 40, 65, 90, and 115~MJy/sr. The yellow contour shows the
3-sigma limit of the SCUBA 850~$\mu m$ emission and therefore the area where
we measured the clump mass. The numbers are the core numbers of Table \ref{core_details}.
The following panels present the VLA and Effelsberg 100m telescope data combined.
The upper right panel shows the integrated emission of the main NH$_3$(1,1)
line in $Jy\, beam^{-1}\, km\, s^{-1}$ from $-41.3$ to $-38.9~km\, s^{-1}$.
The lower left panel presents the peak velocity distribution in $km\, s^{-1}$.
The lower right panel shows the kinetic temperature in K. In each panel white/black
contours present Herschel PACS 70~$\mu m$ emission at levels of 3, 4.5, 6,
and 7.5~mJy/pixel. The NH$_3$(1,1) emission peak is marked in the upper right
panel, and the corresponding spectra are given in Fig.\ref{appendix_spectra_ISOSS22478}.
The synthesized beam and scale are shown in each panel.}
 \label{ISOSS22478_overview}
\end{figure*}

\begin{figure*}
 \centering
 \includegraphics[width=17cm]{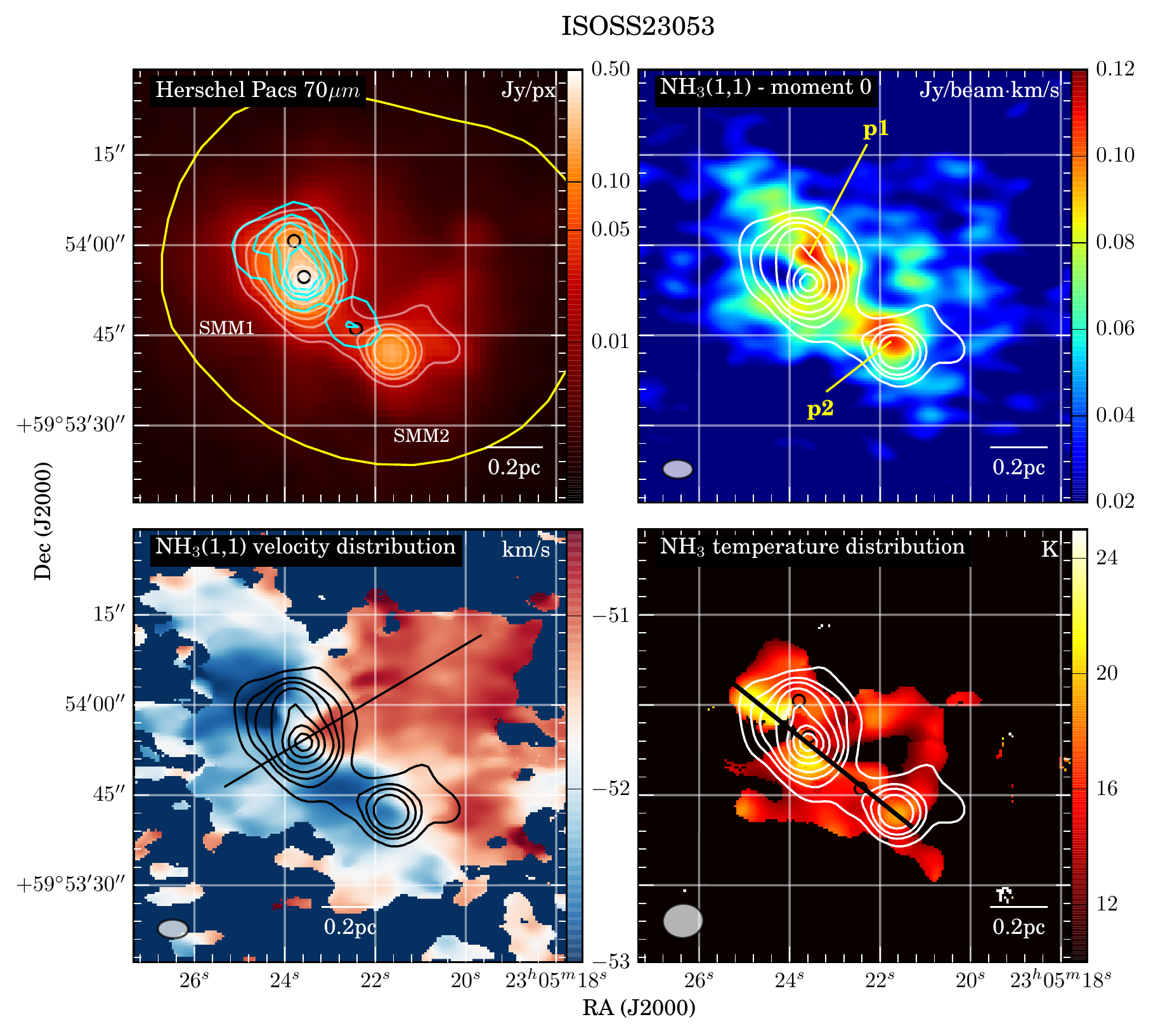}
 \caption{ISOSS23053 - Overview. The upper left panel shows Herschel PACS
70~$\mu m$ emission, and cyan contours present Spitzer MIPS 24~$\mu m$ emission
at levels of 75, 150, 225, and 300~MJy/sr. The yellow contour shows the 3-sigma
limit of the SCUBA 850~$\mu m$ emission and therefore the area where we measured
the clump mass. The following panels present the VLA and Effelsberg 100m
telescope data combined. The upper right panel shows the integrated emission
of the main NH$_3$(1,1) line in $Jy\, beam^{-1}\, km\, s^{-1}$ from $-53.3$
to $-50.1~km\, s^{-1}$. The lower left panel presents the peak velocity distribution
in $km\, s^{-1}$, and the black line indicates the direction of the position-velocity
cut shown in Fig. \ref{ISOSS23053_pvc}. The lower right panel shows the kinetic
temperature in K. We extracted the temperature distribution along the black
line shown in the lower right panel and plotted the values as a function
of distance in Fig. \ref{fig_temp_cut}. The black circles indicate Spitzer
MIPS 24~$\mu m$ emission sources. In each panel white/black contours present
Herschel PACS 70~$\mu m$ emission at levels of 20, 40, 60, 100, 200, 300,
and 400~mJy/pixel. The NH$_3$(1,1) emission peaks are marked in the upper
right panel, and the corresponding spectra are given in Fig.\ref{appendix_spectra_ISOSS23053}.
The synthesized beam and scale are shown in each panel.}
 \label{ISOSS23053_overview}
\end{figure*}

\section{Results}
\label{results}

Figures \ref{IRDC01070_overview} to \ref{ISOSS23053_overview} show the Herschel
PACS 70~$\mu m$ image for each
source in the upper lefthand panel, the integrated NH$_3$(1,1) emission of
the main hyperfine component in the upper righthand corner, the peak velocity
distribution in the lower lefthand panel, and the temperature distribution
in the lower righthand corner.

\subsection{Clump masses}
\label{cloud_masses}
We used the SCUBA 850~$\mu m$ data \citep{Difrancesco2008} to determine clump
masses. These data have a beam size of 22.9'' and an absolute flux uncertainty
of 20\%. The measured flux is used to calculate the clump mass, assuming
optically thin dust emission:

\begin{equation}
M_{\rm{gas}} = \frac{d^2 \; F_{\rm{\nu}} \; R_{\rm{gas}}}{B_{\rm{\nu}}(T) \; \kappa},
\label{mass}
\end{equation} 
where d is the distance of the clump, F$_{\nu}$  the flux density, R$_{gas}$
 the gas-to-dust mass ratio, B$_{\nu}$(T)  the Planck function at temperature
T (estimated from the NH$_3$ observations, see Section \ref{results_temperature}
and Table \ref{sample_details}), and $\kappa$  the dust-mass absorption coefficient.
We assumed a gas-to-dust mass ratio of 100. To determine $\kappa$, we used
the dust model of \citet{Ossenkopf1994} with a density of $10^5~cm^{-3}$
with thin ice mantles and interpolated to 850~$\mu m$, which results in $\kappa
= 1.5~cm^2\:g^{-1}$. The temperature was taken from the Effelsberg 100m telescope
NH$_3$ observations because these observations give an average value for
the temperature. We chose a 3$\sigma$ limit as the boundary for the SCUBA
flux measurements (213~mJy~beam$^{-1}$ and 229~mJy~beam$^{-1}$ for the fundamental
and extended maps, respectively \citep{Difrancesco2008}).\\
The uncertainties in distance (see Table \ref{sample_details}) and temperature
are difficult to estimate, and we assume an uncertainty of $\sim 20\%$. Different
dust models could be applied (thin or thick ice mantles), and this could
change $\kappa$ by $\sim 20\%$. \citet{Difrancesco2008} reports an absolute
flux uncertainty of $20\%$ for the SCUBA 850~\m data. A different gas-to-dust
ratio could also change the mass. Taking all these individual uncertainties
into account reveals an uncertainty in mass of a factor of $\sim$ 4 to 5.\\
Our sample shows a wide spread in clump masses, and the values are given
in Table \ref{sample_details}. The masses given in Table \ref{core_details}
are lower, since these masses only describe the core masses of the infrared
sources, but not the entire clump.

\subsection{Kinematics}
\label{linewidth}
Our limited spectral resolution with the VLA of 0.8~km~s$^{-1}$ prohibits us from resolving very narrow lines. In cases where the line falls within one spectral channel, the CLASS fitting algorithm gives an upper limit for the FWHM of 1.3~km~s$^{-1}$. Several sources in our sample show linewidths that are broader than our resolution limit (e.g., ISOSS23053, ISOSS20153, IRDC079.31), but except for ISOSS20153, further analysis and comparison with the Effelsberg 100m data, which has a high spectral resolution, identified this linewidth broadening as a superposition of several narrow lines that are slightly shifted. The steep velocity gradient within ISOSS23053 is the most prominent example of such a linewidth superposition (see Section \ref{ISOSS23053_velocity_step}). ISOSS20153 is the only source in which the FWHM broadens to $\sim$2.5~km~s$^{-1}$ at the position of an infrared point source. Otherwise, from the combined data, most clouds exhibit linewidths at the upper limit of 1.3~km~s$^{-1}$. On the other hand, the high spectral resolution, but low spatial resolution of our Effelsberg 100m telescope observations indicate an average linewidth for each clump of 1.1 to 1.7~km~s$^{-1}$.\\
With the exception of ISOSS23053, the clumps reveal a smooth peak velocity structure with amplitudes of 1 to 2~km~s$^{-1}$ and projected linear gradients of 5 to 10~km~s$^{-1}$~pc$^{-1}$. In contrast to this, ISOSS23053 reveals two separate velocity components (see Fig. \ref{ISOSS23053_overview}) with a difference of $\sim$1.5~km~s$^{-1}$, but minor velocity fluctuations within each component. The sharp velocity step, situated between the two components, is unique in our sample. Figure \ref{ISOSS23053_pvc} shows a position-velocity cut that is perpendicular to the velocity step, where the velocity gradient increases to at least 30~km~s$^{-1}$~pc$^{-1}$. We discuss the kinematics in general and explore the interpretations of the steep velocity step in Section \ref{discussion_kinematics}.\\

\subsection{Temperature}
\label{results_temperature}
To calculate temperature maps, we used the method described in \citet{Ho_1983}. \citet{Ragan2011} also provides a comprehensive summary of this method. A 3$\sigma$ limit of the NH$_3$(1,1) and (2,2) emission were used for trustworthy results, whereas the NH$_3$(2,2) emission was mainly limiting the area of our temperature maps. To increase the area of reliable NH$_3$(2,2) emission, we used a uvtaper of $30~k\lambda$, which implies giving less weight to long baselines. This method increases the signal-to-noise ratio, but also increases the synthesized beam to $5-7''$. \\
Between 5~K and 25~K, the rotation temperature derived from the NH$_3$(1,1) and (2,2) lines can be converted into kinetic temperatures well \citep{Danby1988,Tafalla2004}. However, above 25~K, these low-level NH$_3$ lines are no longer sensitive to the actual kinetic temperature, so higher temperatures have to be treated with care. In the following, we talk about the kinetic temperature.\\
The observed kinetic temperatures within our sample range from 10 - 30~K and show a clumpy structure down to the resolution limit. Several sources (e.g., IRDC079.31, ISOSS23053 SMM2) show signs of increasing temperatures of $20-30~K$ toward infrared emission sources (see Section \ref{discussion_temperature} for further details). On the other hand, several sources (e.g., ISOSS20153) show marginal ammonia emission close to infrared emission sources (see Section \ref{ammonia_morphology}), making it impossible to measure the temperature in these regions.\\ 

\section{Discussion}
\label{discussion}

The discussion is divided in three parts. First we discuss the kinematical structure of our sample and look for signs of collapse. The second part is devoted to the morphological structure of the ammonia emission and a comparison to infrared emission. Finally we discuss the temperature structure and search for external or internal heating sources.

\subsection{Kinematical structure}
\label{discussion_kinematics}
Infrared dark clouds exhibit significant non-thermal linewidths, even though high-resolution observations suggest that IR-dark regions are relatively more quiescent than the surrounding gas \citep[e.g.,][]{Pineda2010, Hernandez2012}. Therefore we expect a linewidth larger than $\sim 0.3~km~s^{-1}$  for our observations (thermal linewidth), but smaller than $\sim 2~km~s^{-1}$ (quiescent interior of IRDCs). As explained in section \ref{linewidth}, our observations are limited to an upper limit of $1.3~km~s^{-1}$ for the linewidth, and most of our sources show this upper limit. In this section we explore what the dynamics of our gas clumps tell us about the onset of high-mass star formation. 

\subsubsection{Virial parameter}

As outlined in Section \ref{linewidth}, the Effelsberg 100m data allow us to determine the spatially averaged linewidths of our clumps. These linewidths are used to calculate virial masses following \citet{MacLaren1988}:
\begin{equation}
M_{\rm{virial}} = k_2\;R\;\Delta v^2,
\label{virial_mass}
\end{equation} 
where M is the mass in M$_\odot$, R  the radius in pc, and $\Delta v$  the velocity FWHM in km$~s^{-1}$. The factor $k_2$ depends on the density profile and has values of $k_2 = 210$ for $\rho = constant$, $k_2 = 190$ for $\rho \propto 1/r$ and $k_2 = 126$ for $\rho \propto 1/r^2$ \citep{MacLaren1988}. The virial parameter is defined as
\begin{equation}
\alpha = \frac{M_{\rm{virial}}}{M_{\rm{gas}}}.
\label{virial_parameter}
\end{equation} 
The gas mass is determined by the continuum data at 850~$\mu m$ (see Section \ref{cloud_masses} for details). We choose a circle of $40''$ diameter around the peak position given in Table \ref{sample_details}, which corresponds to the Effelsberg 100m telescope beam. Table \ref{virial_mass_table} presents the virial mass, the linear diameter of the $40''$ circle, the gas mass and the virial parameter. The gas masses used here are lower than those given in Table \ref{sample_details}, since we only consider the central area of the clump in order to obtain a consistent comparison between the virial and gas mass.

\begin{table*}
\begin{tabular}{c c c c c c c c}
\hline\hline

Object & $\Delta v$ & Linear diameter & M$_{gas.}$  & M$_{vir}$ $(\rho \propto 1/r)$  & M$_{vir}$ $(\rho \propto 1/r^2)$ & $\alpha$ $(\rho \propto 1/r)$  &$\alpha$ $(\rho \propto 1/r^2)$ \\
 & [$km\, s^{-1}$] & [pc] & [M$_\odot$]& [M$_\odot$] & [M$_\odot$] & & \\
\hline
IRDC079.31 & 1.7 & 0.3 & 70 & 180 & 120& 2.4 & 1.6\\

ISOSS18364 & 1.1 & 0.5 & 90 & 110 & 75 & 1.2 & 0.8\\

ISOSS20153 & 1.3 & 0.2 & 20 & 80 & 50 & 4.0 & 2.6\\

ISOSS22478 & 1.1 & 0.6 & 100 & 150 & 100 & 1.5 & 1.0 \\

ISOSS23053 & 1.2 & 0.8 & 370 & 230 & 150 & 0.6 & 0.4\\
\hline
\end{tabular}
\caption{Virial analysis. The masses are measured within a circle of $40''$ diameter around the peak position given in Table \ref{sample_details}. The corresponding linear diameter is given. $\Delta v$ is measured for the peak spectrum observed with the Effelsberg 100m telescope. Virial masses and virial parameter are given, depending on the density profile of the clump. For ISOSS23053, we chose the linewidth of the red component (see Section \ref{discussion_velocity_step}).}
\label{virial_mass_table}
\end{table*}
As explained in Section \ref{cloud_masses}, the calculated gas masses have an uncertainty of a factor of 4-5. The uncertainty for the virial masses is smaller with a factor of 2-3 and is mainly due to the uncertainty in the distance estimate and the density profile. Another uncertainty is the assumption that the gas mass $M_{gas}$, measured via the dust emission at 850$~\mu m,$ and the virial mass $M_{virial}$, measured via the line width of the NH$_3$(1,1) transition, trace the same gas. However, the spatial structure of the Effelsberg 100m NH$_3$(1,1) observations is similar to the 850$~\mu m$ emission, which implies that the assumption is reasonable.\\ 
The virial parameter is an indicator of gravitational stability, because values lower than one indicate gravitational collapse, whereas values over one are consistent with the regions not being bound. Virial parameters close to one indicate a balance between gravity and thermal or turbulent motions. In this simple analysis, we do not consider additional aspects, such as magnetic fields or external pressure. Our sample has virial parameters between 0.4 and 4.0 with a mean of 1.6. Considering the uncertainties in the masses, we report that the sample as a whole is consistent with an approximate virial balance. This result is consistent with a recent ammonia survey of IRDCs by \citet{Chira2013}. They observed 110 IRDCs with the Effelsberg 100m telescope and found an average virial parameter around one. Similar results have been found by \citet{Wienen2012} and \citet{Dunham2011}. However, our sample does contain individual clumps that have virial parameters clearly in excess of unity, or below it. ISOSS23053 is extremely massive and has an estimated virial parameter of 0.4-0.6. This is suggestive of a clump undergoing large scale collapse, as we discuss in Section \ref{discussion_signature_of_collapse}. In contrast, ISSOSS2015 has the lowest gas mass in the sample, and even at the upper end of our uncertainty estimate it can only be marginally bound.\\

\subsubsection{Rotational energy}
As outlined in section \ref{linewidth}, several of our clumps have smooth velocity gradients, which could be due to rotation of the clump. By assuming solid-body rotation of a sphere with uniform density, we can estimate the rotational energy and compare it with the gravitational energy, using the beta parameter \citep{Stahler2004}:
\begin{equation}
\beta = \frac{E_{\rm{rot}}}{E_{\rm{grav}}} = \frac{\Omega^2 cos(i)^2 \; R^3}{3 \; G \; M},
\label{beta}
\end{equation} 
where $\Omega$ is the angular velocity, R the radius of the clump, G the gravitational constant, and M the mass of the rotating clump. The angular velocity depends on the inclination, which we define to be zero if the angular rotation is perpendicular to the line of sight. As we cannot determine the inclination with our observations, we assume it is edge-on ($i = 0$\degr). We focus our analysis on IRDC079.31 and ISOSS18364, as these show a smooth velocity gradient over the entire clump. The corresponding gradients are $\sim 10~km~s^{-1}~pc^{-1}$ and $\sim 6~km~s^{-1}~pc^{-1}$, for both sources in north-south direction. Once again we use the area corresponding to the Effelsberg beam as in the virial analysis. For these regions, the value of beta are $\beta = 0.37$ and $\beta = 0.48$ for IRDC079.31 and ISOSS18364, respectively. Assuming a more realistic density profile of $\rho \propto 1/r^2$ reduces $\beta$ by a factor of 3 \citep{Ragan2012}. The uncertainty of $\beta$ is within the same order of magnitude as the uncertainty in the mass measurement and therefore a factor of $\sim 4 - 5$.\\
Previous studies have indicated that the rotational energy does not contribute significantly to the overall energy budget of the clumps. For example, \citet{Goodman1993} found values of beta $\sim 0.02$ for cores with size $\sim 0.1~pc$. For a homogeneous comparison with our data, we recalculated the $\beta$ values for the IRDCs studied in \citet{Ragan2012}. They report velocity gradients of $\sim 2~km~s^{-1}~pc^{-1}$ for the IRDCs G005.85-0.23 and G024.05-0.22. The corresponding radii and masses (R $\sim$ 0.25~pc and 0.6~pc, M $\sim$ 500~$M_\odot$ and 2500~$M_\odot$, respectively) result in $\beta$ values of 0.01 and 0.03, which are similar to those reported in \citet{Goodman1993}. The $\beta$ values in our study are a factor of 10 larger.\\
Our measurements are close to the breakup speed of spherical clumps, which corresponds to $\beta$ values greater than $1/3$ \citep{Stahler2004}. This is even true if we consider the uncertainties and use a more realistic centrally peaked density profile. This implies that for our sources, the assumption of circular solid body rotation is a poor description of our data. The observed velocity gradient might instead have a different origin, such as gas flows toward the center. Such flows might also be consistent with the steep velocity steps that we see in the observations.\\ 
\subsubsection{Steep velocity step within ISOSS23053}
\label{discussion_velocity_step}
\label{ISOSS23053_velocity_step}
   \begin{figure}
   \centering
   \includegraphics[width=\hsize]{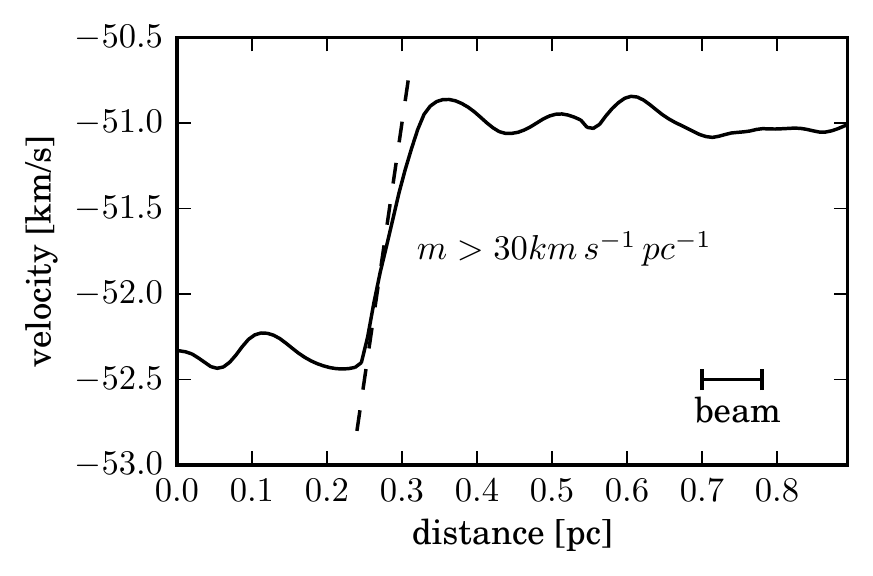}
      \caption{Position velocity cut of ISOSS23053 along the black line indicated in the lower-left panel of Fig. \ref{ISOSS23053_overview}}
         \label{ISOSS23053_pvc}
   \end{figure}
\begin{figure}
  \centering
  \includegraphics[width=\hsize]{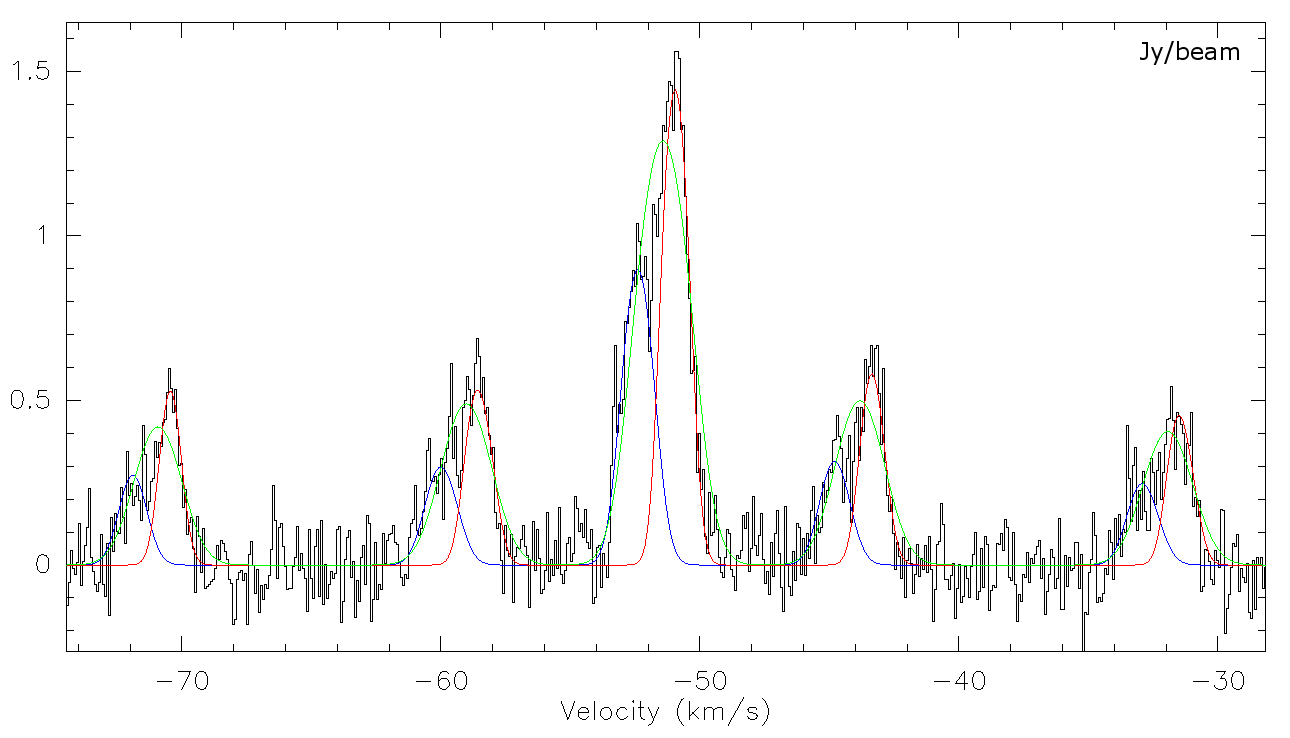}
  \caption{Central spectra of ISOSS23053, as observed with the Effelsberg 100m telescope. The green curve represents one component fit, the blue and red curves represent a two-component fit.}
  \label{ISOSS23053_two_spectra_fit}
\end{figure}
\begin{figure}
 \centering
 \includegraphics[width=\hsize]{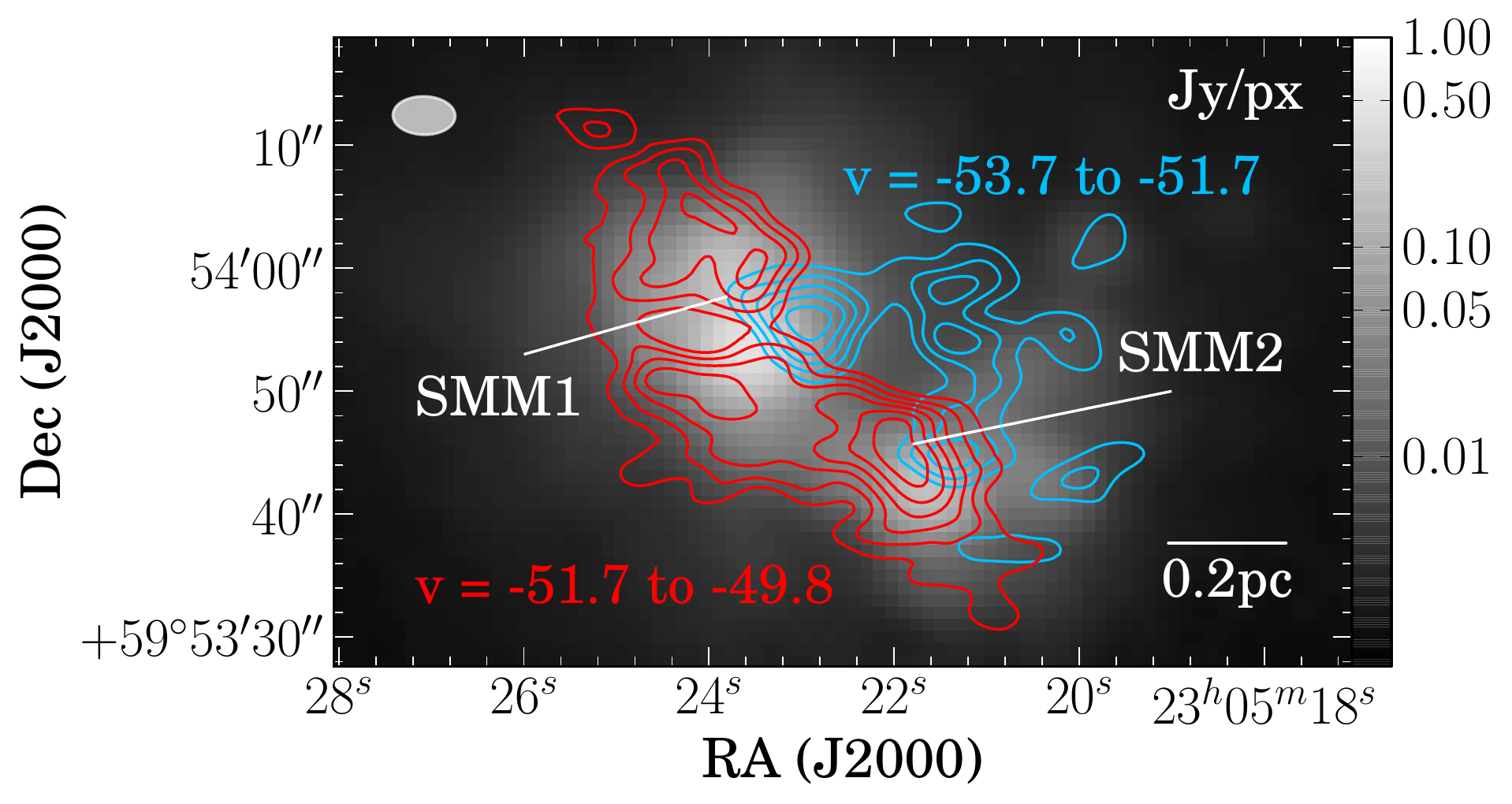}
 \caption{ISOSS23053 - velocity structure. The gray-scale represents Herschel PACS 70~$\mu m$. The red and blue contours represent the NH$_3$ (1,1) moment 0 of each component at levels of 30, 40, 50, 60, 70, and 80~mJy~beam$^{-1}$~km~s$^{-1}$.}
 \label{ISOSS23053_two_components}
\end{figure}

ISOSS23053 is known to be an active star-forming region \citep{Birkmann2007, Pitann2011}. Multiwavelength observations reveal several protostar candidates, outflows, and signs of infall. Submillimeter observations disclose a gas reservoir of several hundred solar masses \citep{Birkmann2007}. However, no previous observations have resolved the kinematic structure spatially.\\
ISOSS23053 harbors two velocity components at $-52.5~km~s^{-1}$ and $-51~km~s^{-1}$.  A spatially unresolved velocity
step is found between these components, which has a gradient larger than $30~km~s^{-1}~pc^{-1}$. The velocity step follows the 24~\m and 70~\m emission from northeast to southwest (see lower left panel in Fig. \ref{ISOSS23053_overview}). Figure \ref{ISOSS23053_pvc} shows the position velocity cut along the axis marked in Fig. \ref{ISOSS23053_overview}.\\
The Effelsberg 100m telescope observations help to define the two velocity components accurately, thanks to their higher spectral resolution (see Fig. \ref{ISOSS23053_two_spectra_fit}). The first component has a velocity range from -53.7 to -51.7~$km~s^{-1}$ (the `blue component'), and the second component is from -51.7 to -49.8~$km~s^{-1}$ (the `red component'). Figure \ref{ISOSS23053_two_components} shows the NH$_3$ (1,1) emission for both components separately. The components overlap at two positions, which are close to SMM1 and SMM2, respectively. The overlap close to SMM1 (northeast) falls within one beam and is therefore not resolved, whereas the second overlap close to SMM2 (southwest) is broader. The Effelsberg 100m telescope observations reveal an average linewidth for each component of $\sim 1.2~km~s^{-1}$ (blue component) and $\sim 0.9~km~^{-1}$ (red component), as illustrated in Fig. \ref{ISOSS23053_two_spectra_fit}.\\
The origin of the velocity step observed in ISOSS23053 is challenging to explain. Because the optical depth is low ($\tau<0.7$), we cannot explain this structure via self absorption of the two components. In the following we propose a solution.\\ 
The observations reveal several velocity components along the line of sight. However, the question is whether these components are spatially connected and/or interact with each other. Figure \ref{ISOSS23053_two_components} shows the ammonia emission of both velocity components separately. The three-dimensional shape of this clump could be a filamentary structure along the line of sight, at which SMM1 shows the beginning and end of the filament, whereas SMM2 is the connection in between. In this picture, the two components at SMM1 could be spatially unconnected.  On the other hand, it could also be that the two velocity components at SMM1 and SMM2 are spatially connected and that the velocity structure stems from the dynamical collapse of converging gas flows. 

This velocity structure is also reported for other sources; for example, \citet{Csengeri2011,Csengeri2011b} report similar velocity steps for DR21(OH). They observed N$_2$H$^+$ with the PdBI and find a velocity shear close to the center, but slightly offset from the strong continuum sources. Their interpretation is that there are converging gas flows toward the central gravitational well. \citet{Beuther2013} also find multiple velocity components in the starless high-mass star-forming region IRDC 18310-4, which are consistent with collapse motions. Recently, \citet{Smith2013} have modeled the velocity structure in dynamically collapsing gas clumps, and they find similar steep velocity gradients close to the clump centers with velocity gradients up to $20~km~s^{-1}~pc^{-1}$. Therefore, ISOSS23053 might undergo global dynamical collapse, and this would explain the active star formation along the velocity step.\\

\subsubsection{Signature of collapse?}
\label{discussion_signature_of_collapse}
To summarize, we find one example in our sample (ISOSS23053) with a steep velocity gradient that might represent ongoing gravitational collapse. In addition we find in two sources (IRDC079.31 and ISOSS18364) smooth velocity gradients, which are poorly described by solid-body rotation and may indicate gas flows toward the clump center. The three other clumps in our sample show no clear evidence of gravitational collapse.\\
After combining these results with recent findings in other sources \citep[e.g.,][]{Csengeri2011,Csengeri2011b,Schneider2010,Beuther2012}, this indicates that large-scale gravitational collapse might be present in several regions of young massive star formation. However, we have not identified any unambiguous signatures yet.

\subsection{Ammonia morphology and correlation with gas and fust}
\label{ammonia_morphology}
Ammonia is a good tracer of cold dense gas \citep[e.g.,][]{Ho_1983,Pillai2006,Tafalla2004,Ragan2011}, but how might this correlation change in the presence of embedded infrared emission sources? Complementary Herschel and Spitzer data allow us to investigate possible deviations from the close correlation between ammonia and submillimeter emission, which indicates dense gas.\\
ISOSS18364 is a clear example of a case where ammonia and submillimeter emission are very closely coupled only in infrared-dark regions. The \citet{Hennemann2009} 3.2\,mm PdBI data show that this IRDC breaks into a north and a south clump (see black contours in Fig. \ref{ISOSS18364_overview}). No infrared sources are evident to the north, and the ammonia peak is consistent with the 3.2\,mm peak. The southern peak is the site of 24 and 70~$\rm{\mu}$m emission, and ammonia does not match the submillimeter emission. This trend is seen consistently throughout the sample. Positions with strong 24~$\rm{\mu}$m emission have little to no ammonia emission (except ISOSS22478 peak 4). On the other hand, positions with strong 70~$\rm{\mu}$m emission and little or no 24~$\rm{\mu}$m emission do show coincident ammonia emission.\\
This observation of decreasing ammonia emission with increasing 24~\m flux is surprising, because other ammonia observations in the literature of similar objects at similar resolutions show a good correlation between 24~\m emission peaks and ammonia emission peaks \citep{Devine2011,Ragan2012}.\\
While we do not have a conclusive explanation for this observation, we do suggest two possibilities in the following.\
The chemistry of ammonia is complicated, and different models show contradictory results. \citet{Miller1997} claims that a low ammonia abundance can be explained by depletion, meaning that ammonia freezes out on grain mantles. More recently, \citet{Vasyunina2012} have reported similar results. They show that the ammonia abundance drops for chemical models that consider gas-grain networks and find that ammonia depletes on grains at typical IRDC conditions ($T=15~K$, $n=10^5~cm^{-3}$) after $10^7$~years. This is a long time scale, but depletion may be faster at the higher densities, which exist around infrared emission sources. Observations also show NH$_3$ ice features in the envelopes of massive young stellar objects \citep{Guertler2002}. However, the explanation of ammonia depletion close to infrared emission sources might be insufficient, because infrared emission sources heat their environment and prevent depletion. Furthermore, chemical models by other authors suggest that ammonia does not deplete, even at high densities \citep{Bergin1997, Nomura2004}.\\
The second possibility is that close to infrared emission sources, a substantial number of ammonia molecules are at higher excited states and therefore `invisible' to our observations of the NH$_3$ (1,1) and (2,2) transition. To investigate this possibility with our observations, we compared the (2,2) emission with the (1,1) emission close to infrared sources. We would expect a higher (2,2) emission close to infrared sources, but this comparison shows no significant enhancement. Furthermore, observations of the NH$_3$(3,3) line of ISOSS18364 and ISOSS22478 with the Effelsberg 100m telescope reveal no detection of this higher excited line (observation noise for ISOSS18364 was $\sigma(T_{MB}) = 49~mK$ (J2000, 18:36:54.7, -2:21:59) and for ISOSS22478 $\sigma(T_{MB}) = 33~mK$ (J2000, 22:47:34.1, +63:56:51), private communication with O. Krause).\\
Both explanations appear insufficient, and we do not have a good explanation for the NH$_3$ emission drop toward several of the detected infrared emission sources.\\

\subsection{Temperature}
\label{discussion_temperature}

  \begin{figure}
   \centering
   \includegraphics[width=\hsize]{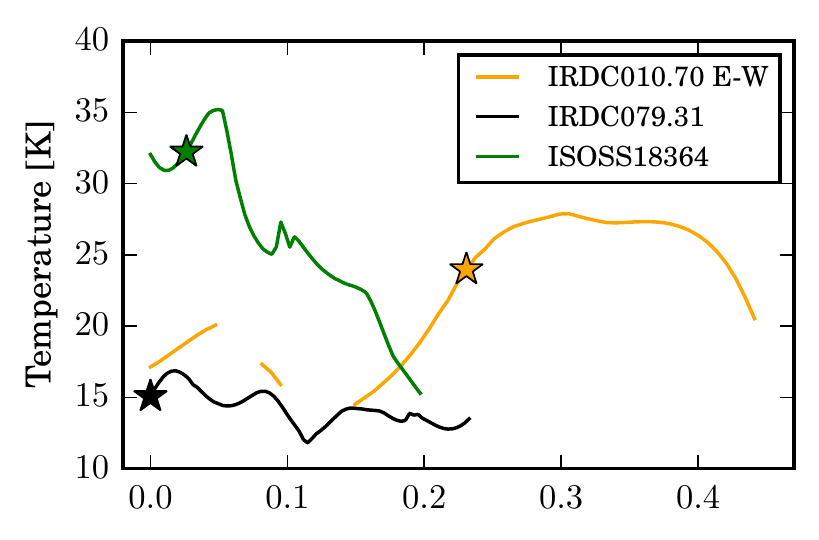}

   \includegraphics[width=\hsize]{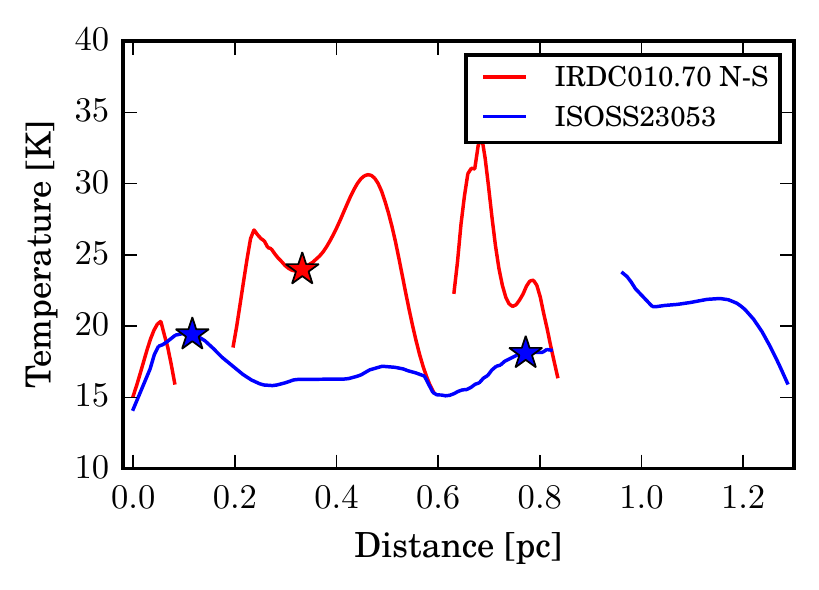}
      \caption{Temperature distribution as a function of the distance along the lines shown in the lower right panels of Figs. \ref{IRDC01070_overview}, \ref{IRDC07931_overview}, \ref{ISOSS18364_overview}, and \ref{ISOSS23053_overview}. In the top panel, the yellow, black, and green lines show the temperature distributions of the sources IRDC010.70 (along east-west), IRDC079.31 and ISOSS18364, respectively. The red and blue lines in the bottom panel show the temperature distributions of the sources IRDC010.70 (along north-south) and ISOSS23053, respectively. The x-axes (distance) are different for the top and bottom panels. The stars mark the positions of corresponding 70~$\mu m$ emission sources.}
         \label{fig_temp_cut}
   \end{figure}

The measured kinetic temperature values in our sample are consistent with similar studies in the literature. Warmer kinetic temperatures are found around active star formation centers \citep[e.g.,][]{Wang2008, Devine2011, Lu2014}, and the coldest kinetic temperatures are found in dark clouds \citep[e.g.,][]{Ragan2011}. Our high-resolution studies are able to probe localized heating from young stars that previous ammonia studies did not resolve \citep[e.g.,][]{Pillai2006, Chira2013}. This also explains why the temperatures in Table \ref{sample_details} are lower than the temperature maps presented in Figs. \ref{IRDC01070_overview}-\ref{ISOSS23053_overview}, because they are determined with the low-resolution Effelsberg 100m data alone.\\
Far-infrared observations from Herschel also allow us to determine dust temperatures. \citet{Ragan2012b} determined the core temperatures for infrared emission sources, and they are given in table \ref{core_details}. These temperatures are on the order of 20~K, which is in good agreement with our ammonia observations.\\
When searching for temperature variations within the clumps, we focused on two aspects. Do we see temperature increases toward infrared emission sources, and do we see shielding of the clumps from the interstellar radiation field.\\
The first aspect is a challenging because we observe lower ammonia emission close to infrared emission sources (see Section \ref{ammonia_morphology}). Therefore it is difficult to measure the temperature at the position of infrared emission sources (e.g., ISOSS20153). To constrain the relation between infrared emission sources and the kinetic temperature, we extracted the temperature along several lines shown in Figs. \ref{IRDC01070_overview}, \ref{IRDC07931_overview}, \ref{ISOSS18364_overview}, and \ref{ISOSS23053_overview} and show the temperature distribution along these lines as a function of distance in Fig. \ref{fig_temp_cut}. We chose the 70~$\mu m$ emission sources to be located along the line, and the location is marked in Fig. \ref{fig_temp_cut}. We find that some sources in our sample show increasing temperatures up to 30~K toward infrared emission sources (e.g., ISOSS23053 MM2, Fig. \ref{fig_temp_cut}). These temperatures are even higher than the core temperatures reported in \citet{Ragan2012b}. However, this could be due to the lower resolution of the Herschel observatory ($\sim$11.4\arcsec for 160~$\mu m$). 

The core temperatures determined in \citet{Ragan2012b} are average values over the entire core, but embedded protostars could heat the center of these cores significantly. Furthermore, the Herschel data probe dust temperatures, whereas the ammonia data probe gas temperatures. Our observations therefore agree with the picture that infrared emission sources imply an embedded protostar(s) that heats its environment. \citet{Wang2008} find similar results toward the northern region of the IRDC G28.34+0.06, which shows increasing temperatures toward the infrared emission source IRS 2 with a peak rotation temperature of 30~K. On the other hand, \citet{Ragan2011} report a quiescent temperature structure without correlating temperature peaks and infrared emission sources, which might be due to the early evolutionary stage of their sample. 

This is also true for some of our sources, since they show no correlation between the temperature peaks and the infrared emission sources, e.g., IRDC079.31 (black line in Fig. \ref{fig_temp_cut}). For this source we see an ammonia temperature peak that is offset to the infrared emission peak. Young embedded protostars, shown as infrared emission sources, are therefore not the only heating source for dense gas traced by ammonia observations. Other mechanisms, such as turbulent motions, can heat the gas as well.\\
The second aspect, shielding of the clumps from the interstellar radiation field, would induce decreasing temperatures toward infrared dark regions. For smaller and less massive objects, \citet{Launhardt2013} report decreasing temperatures toward the center of starless cores from Herschel far-infrared dust observations. They claim that this decrease is due to the shielding of the surrounding core and that only the outer envelope gets heated by the interstellar radiation field \citep[e.g.,][]{Zucconi2001}. In the absence of strong infrared emission sources, this result can also be found for high-mass IRDCs. \citet{Wang2008} report a decreasing temperture of $\sim15~K$ for the more quiescent southern region of G28.34+0.06.\\
In contrast to this, we do not identify any significant temperature decrease toward infrared dark regions within the two IRDCs of our sample (IRDC010.70, IRDC079.31). IRDC010.70 shows a prominent temperature peak within the northern infrared dark region of more than 35~K. IRDC079.31 shows a constant temperature distribution in the northern infrared dark region with a temperature peak on one side up to 17~K (see lower right panel of Fig. \ref{IRDC07931_overview}). This temperature increase may be an indicator of ongoing star formation at that position, which does not show any far-infrared emission yet, or it indicates external heating from the interstellar radiation field.\\

\section{Conclusions}
\label{conclusions}
We observed NH$_3$ (1,1) and (2,2) lines of six IRDCs with the VLA and five out of these six with the Effelsberg 100m telescope. Our goal was to investigate the kinematics and temperatures of these clumps and compare our results with Herschel infrared observations. Our main conclusions follow.\\
\begin{enumerate}
 \item Our ammonia observations show a close spatial correlation with cold gas and dust, which is observed via infrared absorption or submillimeter continuum emission. We confirmed that ammonia is a good tracer of dense gas. But we also found decreasing ammonia emission close to 24~\m emission sources, which has not been observed to date.
 \item We found a temperature range from 10 to 30~K. Temperatures higher than 30~K have to be treated cautiously, because the method based on the low-excitation NH$_3$ lines is not sensitive to higher temperatures. For some sources, we found a correlation between infrared emission sources and the temperature of the surrounding gas, which indicates that the embedded protostar heats up its environment. On the other hand, several infrared emission peaks do not coincide with ammonia temperature peaks. Because we observed a decreasing ammonia emission toward several infrared emission sources, it is difficult to measure the temperature for these infrared emission peaks. Furthermore, we find temperature peaks in the ammonia data that do not coincide with any infrared emission. Other mechanisms, such as turbulent motions or external heating of the interstellar radiation field, can heat the gas.
 \item We report an upper limit of $\sim 1.3~km~s^{-1}$ for the linewidth in our sample. This is the spectral resolution limit of our observation, and we do not see variations larger than this upper limit within our sources.
 \item Half of our sources indicate evidence for gravitational collapse, either in the form of steep velocity steps or smooth velocity gradients, which are not consistent with rotation and therefore might represent gas flows.
 \item The steep velocity gradient observed in ISOSS23053 is unique in our sample. We find two velocity components with a difference of $1.5~km~s^{-1}$ and a gradient larger than $30~km~s^{-1}~pc^{-1}$. Infrared and submillimeter observations show no evidence of different components. It is striking that different indicators of ongoing star formation are located close to the velocity step. As a result, the most likely explanation are signatures from a dynamical collapse and/or converging flows and sheer motions, which may initiate active star formation in this region. This interpretation agrees with recent theoretical models.
\end{enumerate}

In summary, our data constrains the kinematic and thermodynamic initial conditions of star-forming regions in several ways. Although we did not identify collapse signatures in all fields, we do find evidence of gas flows and collapse motions in several clumps. The absence of these signatures in the remaining sources may indicate that they are either more evolved and the signatures have already vanished or that the NH$_3$ observations are not an unambiguous tracer of gas flows and collapse motions. With respect to the temperature structure, we do see temperature increases for some sources toward the more evolved parts of the clumps, indicated by infrared emission sources. On the other hand, we find sources that show no clear correlation between infrared emission sources and ammonia temperature peaks, which indicates that other mechanism, such as turbulent motions or the interstellar radiation field, can heat the gas as well.

\begin{acknowledgements}
Based on observations with the 100-m telescope of the MPIfR (Max-Planck-Institut für Radioastronomie) at Effelsberg.\\ 
The National Radio Astronomy Observatory is a facility of the National Science Foundation operated under cooperative agreement by Associated Universities, Inc.\\
S.E.R. and R.J.S. acknowledge support from the DFG via the SPP 1573 \textit{Physics of the ISM} (grants RA 2158/1-1 and SM321/1-1). R.J.S. also acknowledge support from the SFB 881 \textit{The Milky Way System}.
\end{acknowledgements}

\bibliographystyle{aa}
\bibliography{../references.bib}

\begin{thebibliography}{77}
\expandafter\ifx\csname natexlab\endcsname\relax\def\natexlab#1{#1}\fi

\bibitem[{{Bergin} \& {Langer}(1997)}]{Bergin1997}
{Bergin}, E.~A. \& {Langer}, W.~D. 1997, \apj, 486, 316

\bibitem[{{Bergin} \& {Tafalla}(2007)}]{Bergin2007}
{Bergin}, E.~A. \& {Tafalla}, M. 2007, \araa, 45, 339

\bibitem[{{Beuther} {et~al.}(2007){Beuther}, {Churchwell}, {McKee}, \&
  {Tan}}]{Beuther2007}
{Beuther}, H., {Churchwell}, E.~B., {McKee}, C.~F., \& {Tan}, J.~C. 2007,
  Protostars and Planets V, 165

\bibitem[{{Beuther} \& {Henning}(2009)}]{Beuther2009}
{Beuther}, H. \& {Henning}, T. 2009, \aap, 503, 859

\bibitem[{{Beuther} {et~al.}(2013){Beuther}, {Linz}, {Tackenberg}, {Henning},
  {Krause}, {Ragan}, {Nielbock}, {Launhardt}, {Bihr}, {Schmiedeke}, {Smith}, \&
  {Sakai}}]{Beuther2013}
{Beuther}, H., {Linz}, H., {Tackenberg}, J., {et~al.} 2013, \aap, 553, A115

\bibitem[{{Beuther} \& {Sridharan}(2007)}]{Beuther2007b}
{Beuther}, H. \& {Sridharan}, T.~K. 2007, \apj, 668, 348

\bibitem[{{Beuther} {et~al.}(2012){Beuther}, {Tackenberg}, {Linz}, {Henning},
  {Krause}, {Ragan}, {Nielbock}, {Launhardt}, {Schmiedeke}, {Schuller},
  {Carlhoff}, {Nguyen-Luong}, \& {Sakai}}]{Beuther2012}
{Beuther}, H., {Tackenberg}, J., {Linz}, H., {et~al.} 2012, \aap, 538, A11

\bibitem[{{Birkmann} {et~al.}(2007){Birkmann}, {Krause}, {Hennemann},
  {Henning}, {Steinacker}, \& {Lemke}}]{Birkmann2007}
{Birkmann}, S.~M., {Krause}, O., {Hennemann}, M., {et~al.} 2007, \aap, 474, 883

\bibitem[{{Bogun} {et~al.}(1996){Bogun}, {Lemke}, {Klaas}, {Herbstmeier},
  {Assendorp}, {Richter}, {Laureijs}, {Kessler}, {Burgdorf}, {Schulz}, {Pelz},
  {Beichman}, \& {Rowan-Robinson}}]{Bogun1996}
{Bogun}, S., {Lemke}, D., {Klaas}, U., {et~al.} 1996, \aap, 315, L71

\bibitem[{{Bonnell} {et~al.}(2004){Bonnell}, {Vine}, \& {Bate}}]{Bonnell2004}
{Bonnell}, I.~A., {Vine}, S.~G., \& {Bate}, M.~R. 2004, \mnras, 349, 735

\bibitem[{{Brogan} {et~al.}(2011){Brogan}, {Hunter}, {Cyganowski}, {Friesen},
  {Chandler}, \& {Indebetouw}}]{Brogan2011}
{Brogan}, C.~L., {Hunter}, T.~R., {Cyganowski}, C.~J., {et~al.} 2011, \apjl,
  739, L16

\bibitem[{{Chira} {et~al.}(2013){Chira}, {Beuther}, {Linz}, {Schuller},
  {Walmsley}, {Menten}, \& {Bronfman}}]{Chira2013}
{Chira}, R.-A., {Beuther}, H., {Linz}, H., {et~al.} 2013, \aap, 552, A40

\bibitem[{{Csengeri} {et~al.}(2011{\natexlab{a}}){Csengeri}, {Bontemps},
  {Schneider}, {Motte}, \& {Dib}}]{Csengeri2011}
{Csengeri}, T., {Bontemps}, S., {Schneider}, N., {Motte}, F., \& {Dib}, S.
  2011{\natexlab{a}}, \aap, 527, A135

\bibitem[{{Csengeri} {et~al.}(2011{\natexlab{b}}){Csengeri}, {Bontemps},
  {Schneider}, {Motte}, {Gueth}, \& {Hora}}]{Csengeri2011b}
{Csengeri}, T., {Bontemps}, S., {Schneider}, N., {et~al.} 2011{\natexlab{b}},
  \apjl, 740, L5

\bibitem[{{Danby} {et~al.}(1988){Danby}, {Flower}, {Valiron}, {Schilke}, \&
  {Walmsley}}]{Danby1988}
{Danby}, G., {Flower}, D.~R., {Valiron}, P., {Schilke}, P., \& {Walmsley},
  C.~M. 1988, \mnras, 235, 229

\bibitem[{{de Wit} {et~al.}(2005){de Wit}, {Testi}, {Palla}, \&
  {Zinnecker}}]{Wit2005}
{de Wit}, W.~J., {Testi}, L., {Palla}, F., \& {Zinnecker}, H. 2005, \aap, 437,
  247

\bibitem[{{Devine} {et~al.}(2011){Devine}, {Chandler}, {Brogan}, {Churchwell},
  {Indebetouw}, {Shirley}, \& {Borg}}]{Devine2011}
{Devine}, K.~E., {Chandler}, C.~J., {Brogan}, C., {et~al.} 2011, \apj, 733, 44

\bibitem[{{Di Francesco} {et~al.}(2008){Di Francesco}, {Johnstone}, {Kirk},
  {MacKenzie}, \& {Ledwosinska}}]{Difrancesco2008}
{Di Francesco}, J., {Johnstone}, D., {Kirk}, H., {MacKenzie}, T., \&
  {Ledwosinska}, E. 2008, \apjs, 175, 277

\bibitem[{{Dunham} {et~al.}(2011){Dunham}, {Rosolowsky}, {Evans}, {Cyganowski},
  \& {Urquhart}}]{Dunham2011}
{Dunham}, M.~K., {Rosolowsky}, E., {Evans}, II, N.~J., {Cyganowski}, C., \&
  {Urquhart}, J.~S. 2011, \apj, 741, 110

\bibitem[{{Egan} {et~al.}(1998){Egan}, {Shipman}, {Price}, {Carey}, {Clark}, \&
  {Cohen}}]{Egan1998}
{Egan}, M.~P., {Shipman}, R.~F., {Price}, S.~D., {et~al.} 1998, \apjl, 494,
  L199

\bibitem[{{Evans} {et~al.}(2002){Evans}, {Shirley}, {Mueller}, \&
  {Knez}}]{Evans2002}
{Evans}, N.~J., {Shirley}, Y.~L., {Mueller}, K.~E., \& {Knez}, C. 2002, in
  Astronomical Society of the Pacific Conference Series, Vol. 267, Hot Star
  Workshop III: The Earliest Phases of Massive Star Birth, ed. P.~{Crowther},
  17

\bibitem[{{Goodman} {et~al.}(1993){Goodman}, {Benson}, {Fuller}, \&
  {Myers}}]{Goodman1993}
{Goodman}, A.~A., {Benson}, P.~J., {Fuller}, G.~A., \& {Myers}, P.~C. 1993,
  \apj, 406, 528

\bibitem[{{G{\"u}rtler} {et~al.}(2002){G{\"u}rtler}, {Klaas}, {Henning},
  {{\'A}brah{\'a}m}, {Lemke}, {Schreyer}, \& {Lehmann}}]{Guertler2002}
{G{\"u}rtler}, J., {Klaas}, U., {Henning}, T., {et~al.} 2002, \aap, 390, 1075

\bibitem[{{Gvaramadze} {et~al.}(2012){Gvaramadze}, {Weidner}, {Kroupa}, \&
  {Pflamm-Altenburg}}]{Gvaramadze2012}
{Gvaramadze}, V.~V., {Weidner}, C., {Kroupa}, P., \& {Pflamm-Altenburg}, J.
  2012, \mnras, 424, 3037

\bibitem[{{Hennemann} {et~al.}(2008){Hennemann}, {Birkmann}, {Krause}, \&
  {Lemke}}]{Hennemann2008}
{Hennemann}, M., {Birkmann}, S.~M., {Krause}, O., \& {Lemke}, D. 2008, \aap,
  485, 753

\bibitem[{{Hennemann} {et~al.}(2009){Hennemann}, {Birkmann}, {Krause}, {Lemke},
  {Pavlyuchenkov}, {More}, \& {Henning}}]{Hennemann2009}
{Hennemann}, M., {Birkmann}, S.~M., {Krause}, O., {et~al.} 2009, \apj, 693,
  1379

\bibitem[{{Henning} {et~al.}(2010){Henning}, {Linz}, {Krause}, {Ragan},
  {Beuther}, {Launhardt}, {Nielbock}, \& {Vasyunina}}]{Henning2010}
{Henning}, T., {Linz}, H., {Krause}, O., {et~al.} 2010, \aap, 518, L95

\bibitem[{{Hernandez} {et~al.}(2012){Hernandez}, {Tan}, {Kainulainen},
  {Caselli}, {Butler}, {Jim{\'e}nez-Serra}, \& {Fontani}}]{Hernandez2012}
{Hernandez}, A.~K., {Tan}, J.~C., {Kainulainen}, J., {et~al.} 2012, \apjl, 756,
  L13

\bibitem[{{Ho} \& {Townes}(1983)}]{Ho_1983}
{Ho}, P.~T.~P. \& {Townes}, C.~H. 1983, \araa, 21, 239

\bibitem[{{Kauffmann} {et~al.}(2013){Kauffmann}, {Pillai}, \&
  {Zhang}}]{Kauffmann2013}
{Kauffmann}, J., {Pillai}, T., \& {Zhang}, Q. 2013, \apjl, 765, L35

\bibitem[{{Klein} {et~al.}(2005){Klein}, {Posselt}, {Schreyer}, {Forbrich}, \&
  {Henning}}]{Klein2005}
{Klein}, R., {Posselt}, B., {Schreyer}, K., {Forbrich}, J., \& {Henning}, T.
  2005, \apjs, 161, 361

\bibitem[{{Klessen}(2011)}]{Klessen2011}
{Klessen}, R.~S. 2011, in EAS Publications Series, Vol.~51, EAS Publications
  Series, ed. {C.~Charbonnel \& T.~Montmerle}, 133--167

\bibitem[{{Klessen} {et~al.}(2005){Klessen}, {Ballesteros-Paredes},
  {V{\'a}zquez-Semadeni}, \& {Dur{\'a}n-Rojas}}]{Klessen2005}
{Klessen}, R.~S., {Ballesteros-Paredes}, J., {V{\'a}zquez-Semadeni}, E., \&
  {Dur{\'a}n-Rojas}, C. 2005, \apj, 620, 786

\bibitem[{{Krause} {et~al.}(2003){Krause}, {Lemke}, {Vavrek}, {T{\'o}th},
  {Stickel}, \& {Klaas}}]{Krause2003}
{Krause}, O., {Lemke}, D., {Vavrek}, R., {et~al.} 2003, in Astronomical Society
  of the Pacific Conference Series, Vol. 287, Galactic Star Formation Across
  the Stellar Mass Spectrum, ed. J.~M. {De Buizer} \& N.~S. {van der Bliek},
  174--179

\bibitem[{{Krause} {et~al.}(2004){Krause}, {Vavrek}, {Birkmann}, {Klaas},
  {Stickel}, {T{\'o}th}, \& {Lemke}}]{Krause2004}
{Krause}, O., {Vavrek}, R., {Birkmann}, S., {et~al.} 2004, Baltic Astronomy,
  13, 407

\bibitem[{{Lada} \& {Lada}(2003)}]{Lada2003}
{Lada}, C.~J. \& {Lada}, E.~A. 2003, \araa, 41, 57

\bibitem[{{Launhardt} {et~al.}(2013){Launhardt}, {Stutz}, {Schmiedeke},
  {Henning}, {Krause}, {Balog}, {Beuther}, {Birkmann}, {Hennemann},
  {Kainulainen}, {Khanzadyan}, {Linz}, {Lippok}, {Nielbock}, {Pitann}, {Ragan},
  {Risacher}, {Schmalzl}, {Shirley}, {Stecklum}, {Steinacker}, \&
  {Tackenberg}}]{Launhardt2013}
{Launhardt}, R., {Stutz}, A.~M., {Schmiedeke}, A., {et~al.} 2013, \aap, 551,
  A98

\bibitem[{{Lu} {et~al.}(2014){Lu}, {Zhang}, {Liu}, {Wang}, \& {Gu}}]{Lu2014}
{Lu}, X., {Zhang}, Q., {Liu}, H.~B., {Wang}, J., \& {Gu}, Q. 2014, \apj, 790,
  84

\bibitem[{{MacLaren} {et~al.}(1988){MacLaren}, {Richardson}, \&
  {Wolfendale}}]{MacLaren1988}
{MacLaren}, I., {Richardson}, K.~M., \& {Wolfendale}, A.~W. 1988, \apj, 333,
  821

\bibitem[{{Millar} {et~al.}(1997){Millar}, {MacDonald}, \& {Gibb}}]{Miller1997}
{Millar}, T.~J., {MacDonald}, G.~H., \& {Gibb}, A.~G. 1997, \aap, 325, 1163

\bibitem[{{Nomura} \& {Millar}(2004)}]{Nomura2004}
{Nomura}, H. \& {Millar}, T.~J. 2004, \aap, 414, 409

\bibitem[{{Ossenkopf} \& {Henning}(1994)}]{Ossenkopf1994}
{Ossenkopf}, V. \& {Henning}, T. 1994, \aap, 291, 943

\bibitem[{{Perault} {et~al.}(1996){Perault}, {Omont}, {Simon}, {Seguin},
  {Ojha}, {Blommaert}, {Felli}, {Gilmore}, {Guglielmo}, {Habing}, {Price},
  {Robin}, {de Batz}, {Cesarsky}, {Elbaz}, {Epchtein}, {Fouque}, {Guest},
  {Levine}, {Pollock}, {Prusti}, {Siebenmorgen}, {Testi}, \&
  {Tiphene}}]{Perault1996}
{Perault}, M., {Omont}, A., {Simon}, G., {et~al.} 1996, \aap, 315, L165

\bibitem[{{Peretto} \& {Fuller}(2009)}]{Peretto2009}
{Peretto}, N. \& {Fuller}, G.~A. 2009, \aap, 505, 405

\bibitem[{{Pilbratt} {et~al.}(2010){Pilbratt}, {Riedinger}, {Passvogel},
  {Crone}, {Doyle}, {Gageur}, {Heras}, {Jewell}, {Metcalfe}, {Ott}, \&
  {Schmidt}}]{Pilbratt2010}
{Pilbratt}, G.~L., {Riedinger}, J.~R., {Passvogel}, T., {et~al.} 2010, \aap,
  518, L1

\bibitem[{{Pillai} {et~al.}(2006{\natexlab{a}}){Pillai}, {Wyrowski}, {Carey},
  \& {Menten}}]{Pillai2006}
{Pillai}, T., {Wyrowski}, F., {Carey}, S.~J., \& {Menten}, K.~M.
  2006{\natexlab{a}}, \aap, 450, 569

\bibitem[{{Pillai} {et~al.}(2006{\natexlab{b}}){Pillai}, {Wyrowski}, {Menten},
  \& {Kr{\"u}gel}}]{Pillai2006b}
{Pillai}, T., {Wyrowski}, F., {Menten}, K.~M., \& {Kr{\"u}gel}, E.
  2006{\natexlab{b}}, \aap, 447, 929

\bibitem[{{Pineda} {et~al.}(2010){Pineda}, {Goodman}, {Arce}, {Caselli},
  {Foster}, {Myers}, \& {Rosolowsky}}]{Pineda2010}
{Pineda}, J.~E., {Goodman}, A.~A., {Arce}, H.~G., {et~al.} 2010, \apjl, 712,
  L116

\bibitem[{{Pitann} {et~al.}(2011){Pitann}, {Hennemann}, {Birkmann}, {Bouwman},
  {Krause}, \& {Henning}}]{Pitann2011}
{Pitann}, J., {Hennemann}, M., {Birkmann}, S., {et~al.} 2011, \apj, 743, 93

\bibitem[{{Purcell} {et~al.}(2012){Purcell}, {Longmore}, {Walsh}, {Whiting},
  {Breen}, {Britton}, {Brooks}, {Burton}, {Cunningham}, {Green},
  {Harvey-Smith}, {Hindson}, {Hoare}, {Indermuehle}, {Jones}, {Lo}, {Lowe},
  {Phillips}, {Thompson}, {Urquhart}, {Voronkov}, \& {White}}]{Purcell2012}
{Purcell}, C.~R., {Longmore}, S.~N., {Walsh}, A.~J., {et~al.} 2012, \mnras,
  426, 1972

\bibitem[{{Ragan} {et~al.}(2012{\natexlab{a}}){Ragan}, {Henning}, {Krause},
  {Pitann}, {Beuther}, {Linz}, {Tackenberg}, {Balog}, {Hennemann}, {Launhardt},
  {Lippok}, {Nielbock}, {Schmiedeke}, {Schuller}, {Steinacker}, {Stutz}, \&
  {Vasyunina}}]{Ragan2012b}
{Ragan}, S., {Henning}, T., {Krause}, O., {et~al.} 2012{\natexlab{a}}, \aap,
  547, A49

\bibitem[{{Ragan} {et~al.}(2009){Ragan}, {Bergin}, \& {Gutermuth}}]{Ragan2009}
{Ragan}, S.~E., {Bergin}, E.~A., \& {Gutermuth}, R.~A. 2009, \apj, 698, 324

\bibitem[{{Ragan} {et~al.}(2011){Ragan}, {Bergin}, \& {Wilner}}]{Ragan2011}
{Ragan}, S.~E., {Bergin}, E.~A., \& {Wilner}, D. 2011, \apj, 736, 163

\bibitem[{{Ragan} {et~al.}(2012{\natexlab{b}}){Ragan}, {Heitsch}, {Bergin}, \&
  {Wilner}}]{Ragan2012}
{Ragan}, S.~E., {Heitsch}, F., {Bergin}, E.~A., \& {Wilner}, D.
  2012{\natexlab{b}}, \apj, 746, 174

\bibitem[{{Rathborne} {et~al.}(2010){Rathborne}, {Jackson}, {Chambers},
  {Stojimirovic}, {Simon}, {Shipman}, \& {Frieswijk}}]{Rathborne2010}
{Rathborne}, J.~M., {Jackson}, J.~M., {Chambers}, E.~T., {et~al.} 2010, \apj,
  715, 310

\bibitem[{{Rathborne} {et~al.}(2008){Rathborne}, {Jackson}, {Zhang}, \&
  {Simon}}]{Rathborne2008}
{Rathborne}, J.~M., {Jackson}, J.~M., {Zhang}, Q., \& {Simon}, R. 2008, \apj,
  689, 1141

\bibitem[{{Rathborne} {et~al.}(2007){Rathborne}, {Simon}, \&
  {Jackson}}]{Rathborne2007}
{Rathborne}, J.~M., {Simon}, R., \& {Jackson}, J.~M. 2007, \apj, 662, 1082

\bibitem[{{Redman} {et~al.}(2003){Redman}, {Feldman}, {Wyrowski},
  {C{\^o}t{\'e}}, {Carey}, \& {Egan}}]{Redman2003}
{Redman}, R.~O., {Feldman}, P.~A., {Wyrowski}, F., {et~al.} 2003, \apj, 586,
  1127

\bibitem[{{Reid} {et~al.}(2009){Reid}, {Menten}, {Zheng}, {Brunthaler},
  {Moscadelli}, {Xu}, {Zhang}, {Sato}, {Honma}, {Hirota}, {Hachisuka}, {Choi},
  {Moellenbrock}, \& {Bartkiewicz}}]{Reid2009}
{Reid}, M.~J., {Menten}, K.~M., {Zheng}, X.~W., {et~al.} 2009, \apj, 700, 137

\bibitem[{{S{\'a}nchez-Monge} {et~al.}(2013){S{\'a}nchez-Monge}, {Palau},
  {Fontani}, {Busquet}, {Ju{\'a}rez}, {Estalella}, {Tan}, {Sep{\'u}lveda},
  {Ho}, {Zhang}, \& {Kurtz}}]{Sanchez2013}
{S{\'a}nchez-Monge}, {\'A}., {Palau}, A., {Fontani}, F., {et~al.} 2013, \mnras,
  432, 3288

\bibitem[{{Schneider} {et~al.}(2010){Schneider}, {Csengeri}, {Bontemps},
  {Motte}, {Simon}, {Hennebelle}, {Federrath}, \& {Klessen}}]{Schneider2010}
{Schneider}, N., {Csengeri}, T., {Bontemps}, S., {et~al.} 2010, \aap, 520, A49

\bibitem[{{Simon} {et~al.}(2006){Simon}, {Rathborne}, {Shah}, {Jackson}, \&
  {Chambers}}]{Simon2006}
{Simon}, R., {Rathborne}, J.~M., {Shah}, R.~Y., {Jackson}, J.~M., \&
  {Chambers}, E.~T. 2006, \apj, 653, 1325

\bibitem[{{Smith} {et~al.}(2013){Smith}, {Shetty}, {Beuther}, {Klessen}, \&
  {Bonnell}}]{Smith2013}
{Smith}, R.~J., {Shetty}, R., {Beuther}, H., {Klessen}, R.~S., \& {Bonnell},
  I.~A. 2013, \apj, 771, 24

\bibitem[{Stahler {et~al.}(2005)Stahler, Palla, \& Palla}]{Stahler2004}
Stahler, S.~W., Palla, F., \& Palla, F. 2005, The Formation of Stars (Physics
  Textbook), 1st edn. (Wiley-VCH)

\bibitem[{{Tafalla} {et~al.}(2004){Tafalla}, {Myers}, {Caselli}, \&
  {Walmsley}}]{Tafalla2004}
{Tafalla}, M., {Myers}, P.~C., {Caselli}, P., \& {Walmsley}, C.~M. 2004, \aap,
  416, 191

\bibitem[{{Tan} {et~al.}(2014){Tan}, {Beltr{\'a}n}, {Caselli}, {Fontani},
  {Fuente}, {Krumholz}, {McKee}, \& {Stolte}}]{Tan2014}
{Tan}, J.~C., {Beltr{\'a}n}, M.~T., {Caselli}, P., {et~al.} 2014, Protostars
  and Planets VI, 149

\bibitem[{{Vasyunina} {et~al.}(2009){Vasyunina}, {Linz}, {Henning}, {Stecklum},
  {Klose}, \& {Nyman}}]{Vasyunina2009}
{Vasyunina}, T., {Linz}, H., {Henning}, T., {et~al.} 2009, \aap, 499, 149

\bibitem[{{Vasyunina} {et~al.}(2012){Vasyunina}, {Vasyunin}, {Herbst}, \&
  {Linz}}]{Vasyunina2012}
{Vasyunina}, T., {Vasyunin}, A.~I., {Herbst}, E., \& {Linz}, H. 2012, \apj,
  751, 105

\bibitem[{{Wang} {et~al.}(2008){Wang}, {Zhang}, {Pillai}, {Wyrowski}, \&
  {Wu}}]{Wang2008}
{Wang}, Y., {Zhang}, Q., {Pillai}, T., {Wyrowski}, F., \& {Wu}, Y. 2008, \apjl,
  672, L33

\bibitem[{{Wienen} {et~al.}(2012){Wienen}, {Wyrowski}, {Schuller}, {Menten},
  {Walmsley}, {Bronfman}, \& {Motte}}]{Wienen2012}
{Wienen}, M., {Wyrowski}, F., {Schuller}, F., {et~al.} 2012, \aap, 544, A146

\bibitem[{{Williams} {et~al.}(2000){Williams}, {Blitz}, \&
  {McKee}}]{Williams2000}
{Williams}, J.~P., {Blitz}, L., \& {McKee}, C.~F. 2000, Protostars and Planets
  IV, 97

\bibitem[{{Wyrowski}(2008)}]{Wyrowski2008}
{Wyrowski}, F. 2008, in Astronomical Society of the Pacific Conference Series,
  Vol. 387, Massive Star Formation: Observations Confront Theory, ed.
  H.~{Beuther}, H.~{Linz}, \& T.~{Henning}, 3

\bibitem[{{Zhang} \& {Wang}(2011)}]{Zhang2011}
{Zhang}, Q. \& {Wang}, K. 2011, \apj, 733, 26

\bibitem[{{Zhang} {et~al.}(2009){Zhang}, {Wang}, {Pillai}, \&
  {Rathborne}}]{Zhang2009}
{Zhang}, Q., {Wang}, Y., {Pillai}, T., \& {Rathborne}, J. 2009, \apj, 696, 268

\bibitem[{{Zijlstra} {et~al.}(2008){Zijlstra}, {van Hoof}, \&
  {Perley}}]{Zijlstra2008}
{Zijlstra}, A.~A., {van Hoof}, P.~A.~M., \& {Perley}, R.~A. 2008, \apj, 681,
  1296

\bibitem[{{Zinnecker} \& {Yorke}(2007)}]{Zinnecker2007}
{Zinnecker}, H. \& {Yorke}, H.~W. 2007, \araa, 45, 481

\bibitem[{{Zucconi} {et~al.}(2001){Zucconi}, {Walmsley}, \&
  {Galli}}]{Zucconi2001}
{Zucconi}, A., {Walmsley}, C.~M., \& {Galli}, D. 2001, \aap, 376, 650

\end{thebibliography}

\begin{appendix}
\section{Ammonia spectra}
\begin{table*}
\centering                
\begin{tabular}{  l  c  c  c  c  c c  }
\hline\hline
        Object & $\alpha$ (J2000) & $\delta$ (J2000) & $T_{ant} (1,1)$ & $T_{ant} (2,2)$ & $v_{peak}$ & $\tau_{peak}$ \\ 
        Name  & [h:m:s] & $[^{\circ} : ' : '']$ & [$Jy\,beam^{-1}$] & [$Jy\,beam^{-1}$] & $[km\,s^{-1}]$ &  \\   
   \hline
        IRDC010.70 p1 & 18:09:45.589 & -19:42:08.800 & 0.056 & 0.038 & 28.8 & 3.4 \\ 
        IRDC079.31 100m & 20:31:57.800 & 40:18:19.900 & 1.49 & 0.43 & 1.59 & 3.0 \\ 
        IRDC079.31 p1 & 20:31:57.860 & 40:18:31.613 & 0.117 & 0.045 & 1.8 & 2.9 \\ 
        IRDC079.31 p2 & 20:31:57.545 & 40:18:21.213 & 0.106 & 0.034 & 1.6 & 3.6 \\ 
        ISOSS18364 100m & 18:36:36.000 & -02:21:44.700 & 0.46 & - & 35.1 & 2.51 \  \\ 
        ISOSS18364 p1 & 18:36:35.735 & -02:21:40.181 & 0.052 & 0.014 & 35.1 & 0.1 \\ 
        ISOSS20153 100m & 20:15:21.700 & 34:53:44.900 & 0.39 & 0.13 & 2.4 & 1.3 \\ 
        ISOSS20153 p1 & 20:15:21.928 & 34:53:28.807 & 0.086 & 0.016 & 2.1 & 0.4 \\ 
        ISOSS20153 p2 & 20:15:21.311 & 34:53:40.001 & 0.066 & - & 2.1 & 0.1 \  \\ 
        ISOSS22478 100m & 22:47:49.700 & 63:56:45.500 & 0.54 & - & -40.0 & 1.5  \  \\ 
        ISOSS22478 p1 & 22:47:50.027 & 63:56:47.414 & 0.068 & - & -40.4 & 0.3  \  \\ 
        ISOSS23053 100m & 23:05:21.900 & 59:53:50.000 & 1.53 & 0.62 & -51.6 & 0.9 \\
        ISOSS23053 p1 & 23:05:23.563 & 59:53:58.409 & 0.070 & 0.022 & -51.9 & 0.72 \\ 
        ISOSS23053 p2 & 23:05:21.755 & 59:53:44.016 & 0.110 & 0.031 & -52.3 & 0.47 \\ \hline
\end{tabular}
\caption{Summary of ammonia emission peaks shown in Fig. \ref{IRDC01070_overview}-\ref{ISOSS23053_overview}. The position and fit parameter for the ammonia emission peaks as well as for the ammonia observations with the Effelsberg 100m are given. The corresponding spectra are shown in Fig.\ref{appendix_spectra_IRDC01070}-\ref{appendix_spectra_ISOSS23053}. As the VLA observations do not resolve the line width, we do not mention it here. The line widths of the Effelsberg 100m observations are given in Table \ref{virial_mass_table}.}
\label{table_ammonia_spectra}
\end{table*}

\begin{figure*}
    \centering
       \includegraphics[width=17cm]{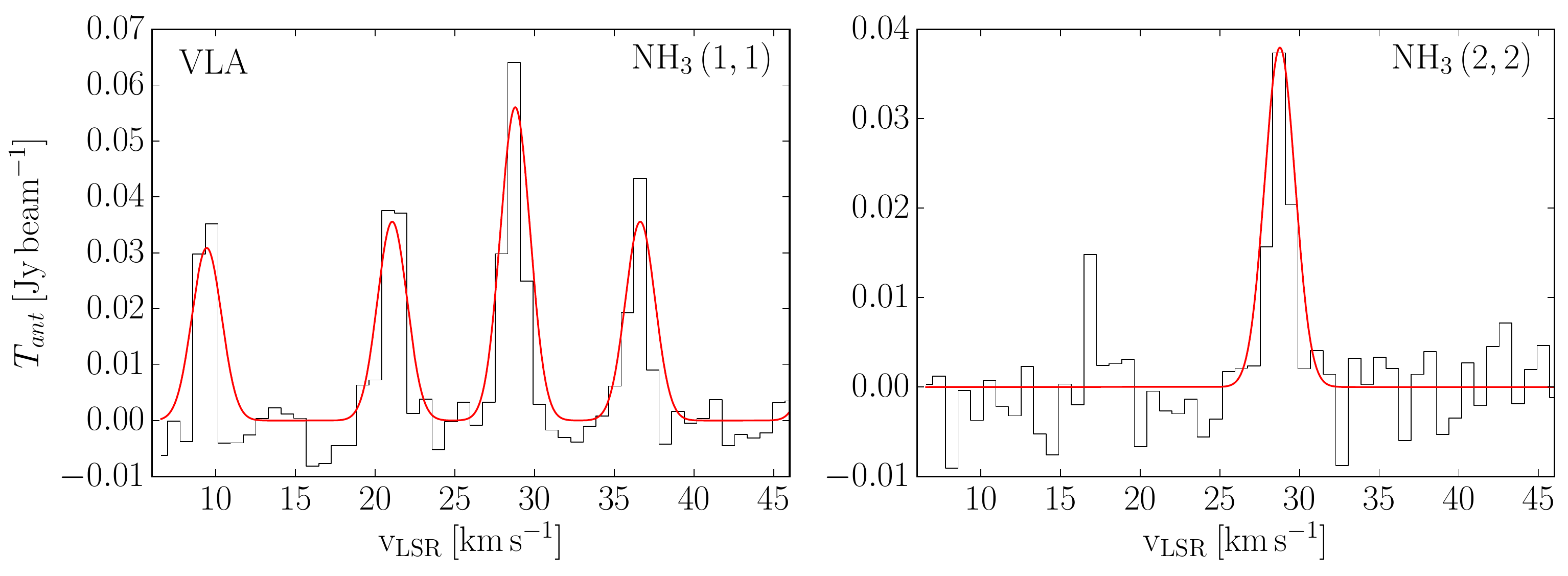}
       \caption{The VLA NH$_3$ spectra of IRDC010.70 at the position p1. The left-hand side shows the NH$_3$(1,1) spectra and the right-hand side shows the NH$_3$(2,2) spectra. The red line corresponds to the fit. As we do not observe the hyperfine structure of the NH$_3$(2,2) transition, we show a Gaussian fit to the main component. The fit parameters and coordinates of the spectra are given in Table \ref{table_ammonia_spectra}.}
        \label{appendix_spectra_IRDC01070}
\end{figure*}

\begin{figure*}
    \centering
       \includegraphics[width=17cm]{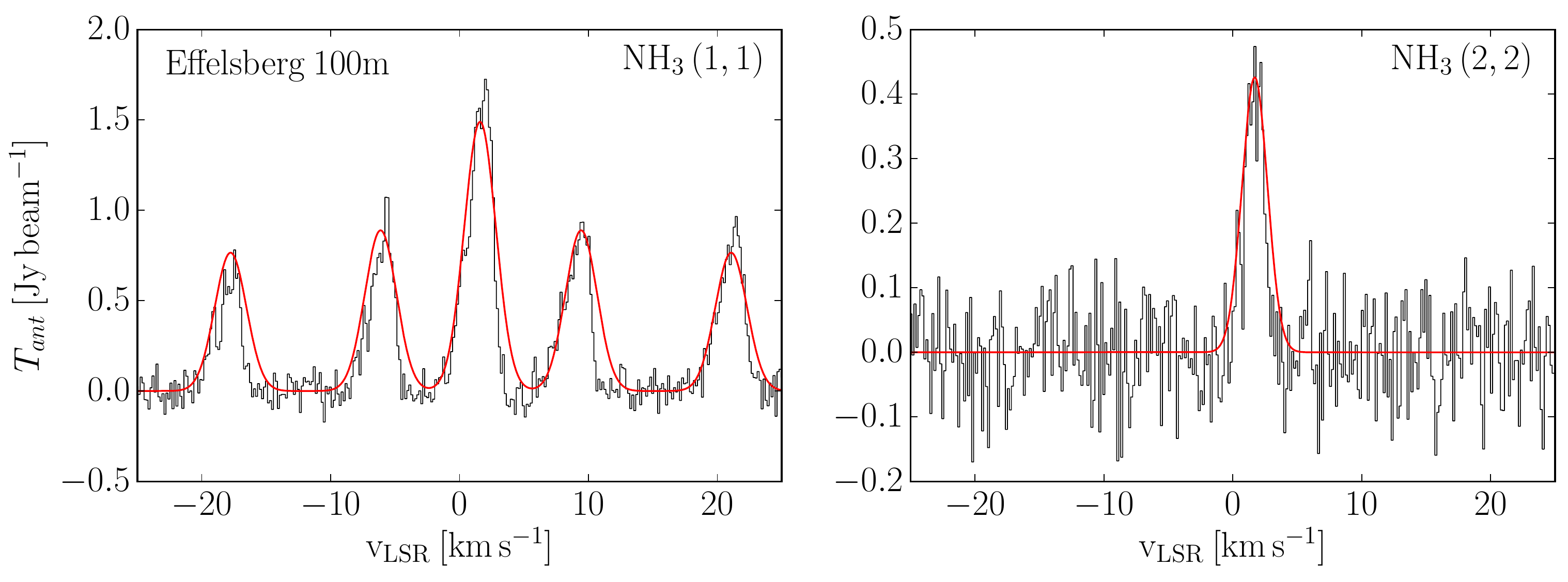}

       \includegraphics[width=17cm]{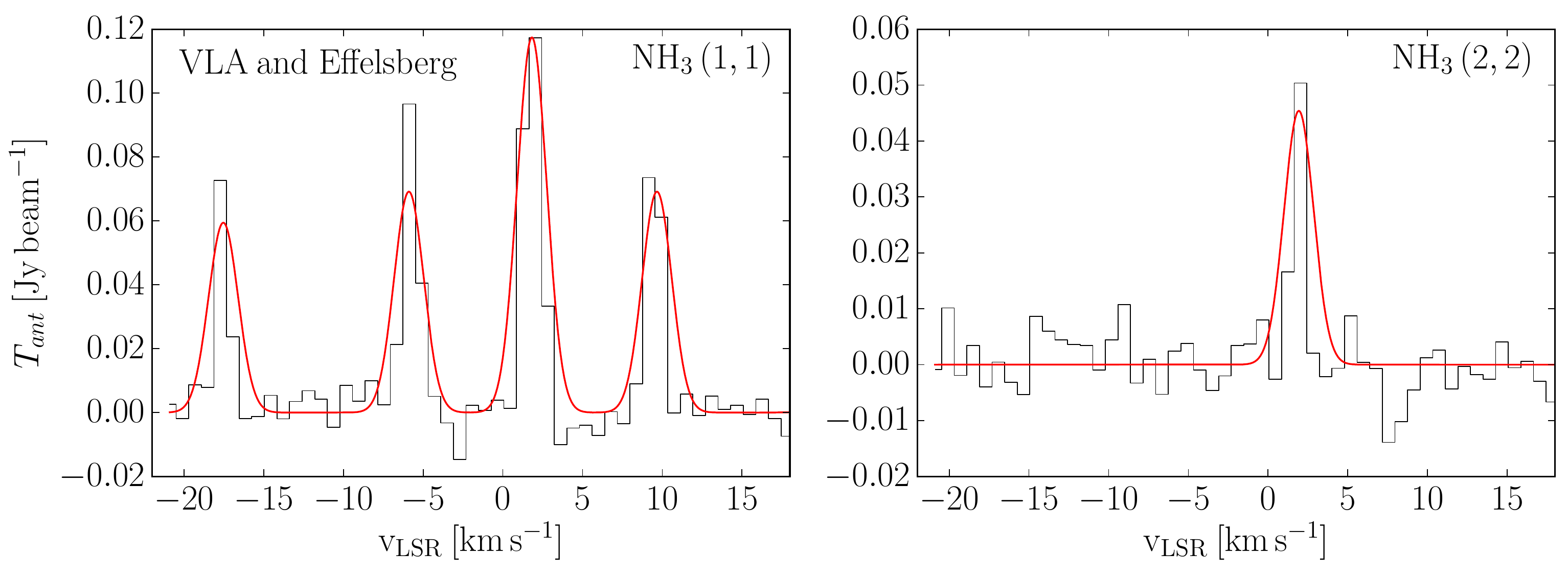}

       \includegraphics[width=17cm]{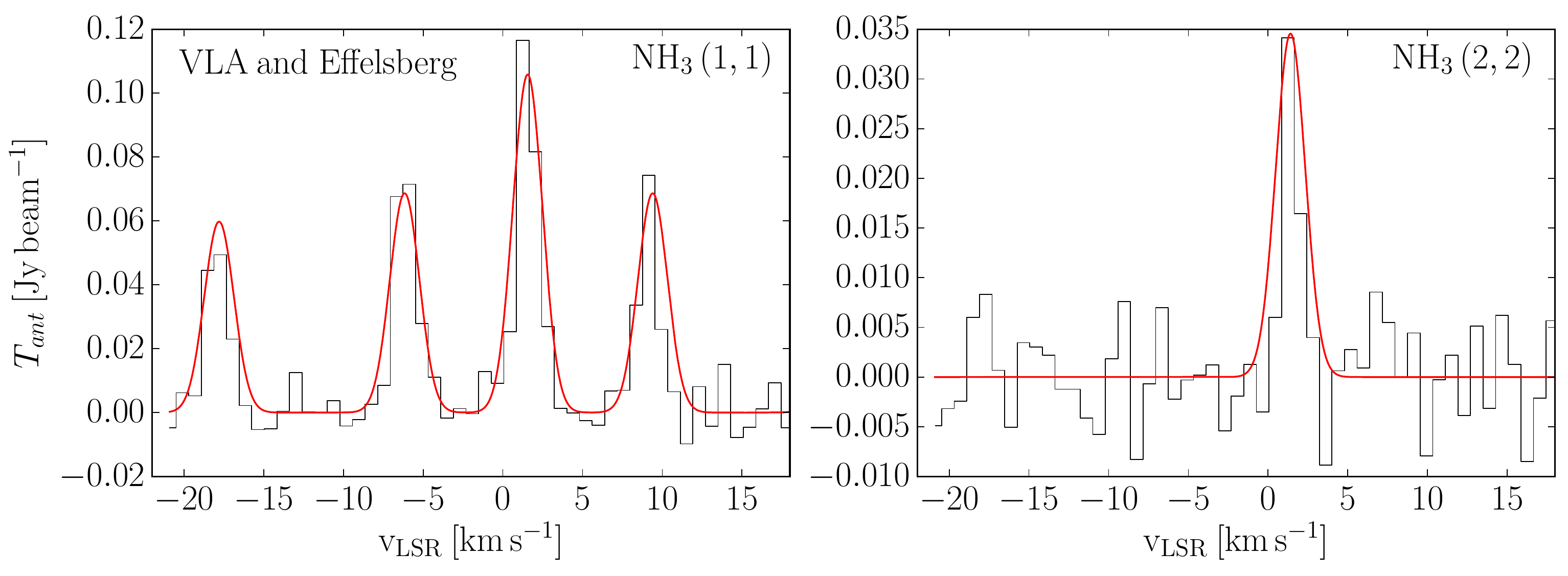}
       \caption{The NH$_3$ spectra of IRDC079.31. The left-hand side shows the NH$_3$(1,1) spectra and the right-hand side shows the NH$_3$(2,2) spectra. The top two panels show the Effelsberg 100m observations, the central panels show the combined VLA and Effelsberg 100m spectra at the position p1 and the bottom panels show the combined VLA and Effelsberg 100m spectra at the position p2. The red line corresponds to the fit. As we do not observe the hyperfine structure of the NH$_3$(2,2) transition, we show a Gaussian fit to the main component. The fit parameters and coordinates of the spectra are given in Table \ref{table_ammonia_spectra}.}
        \label{appendix_spectra_IRDC079.31}
\end{figure*}

\begin{figure*}
    \centering
       \includegraphics[width=17cm]{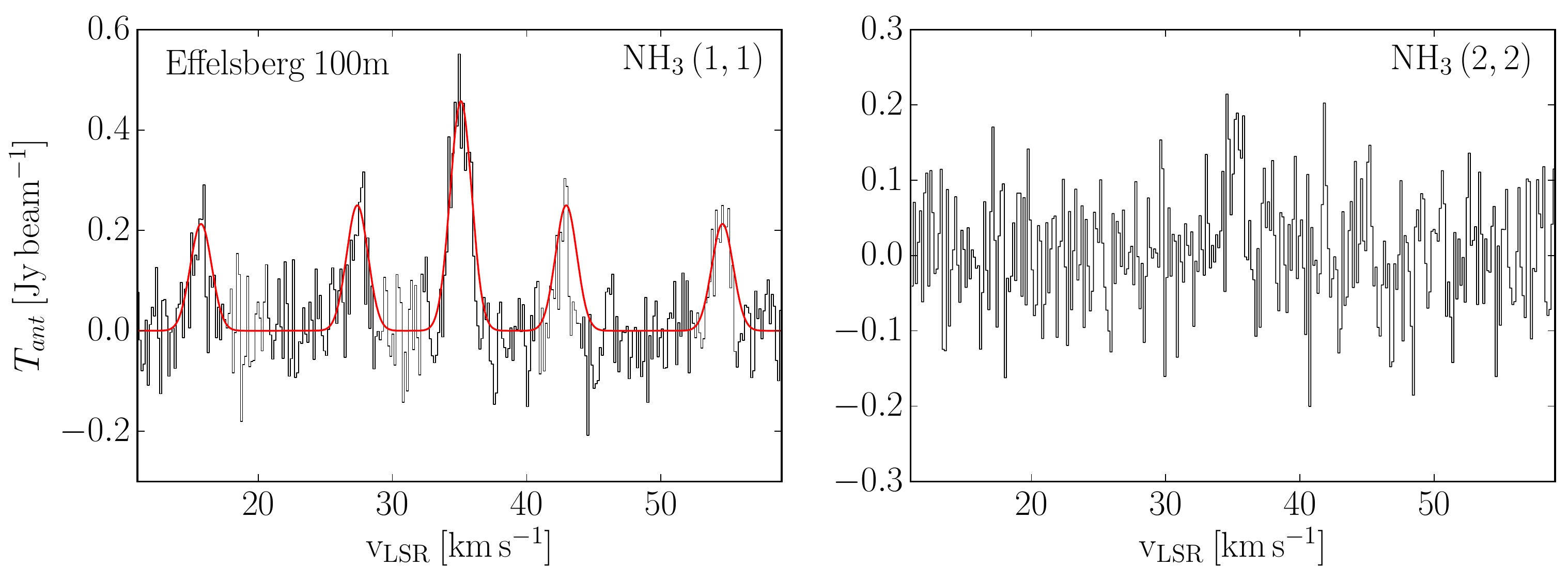}

       \includegraphics[width=17cm]{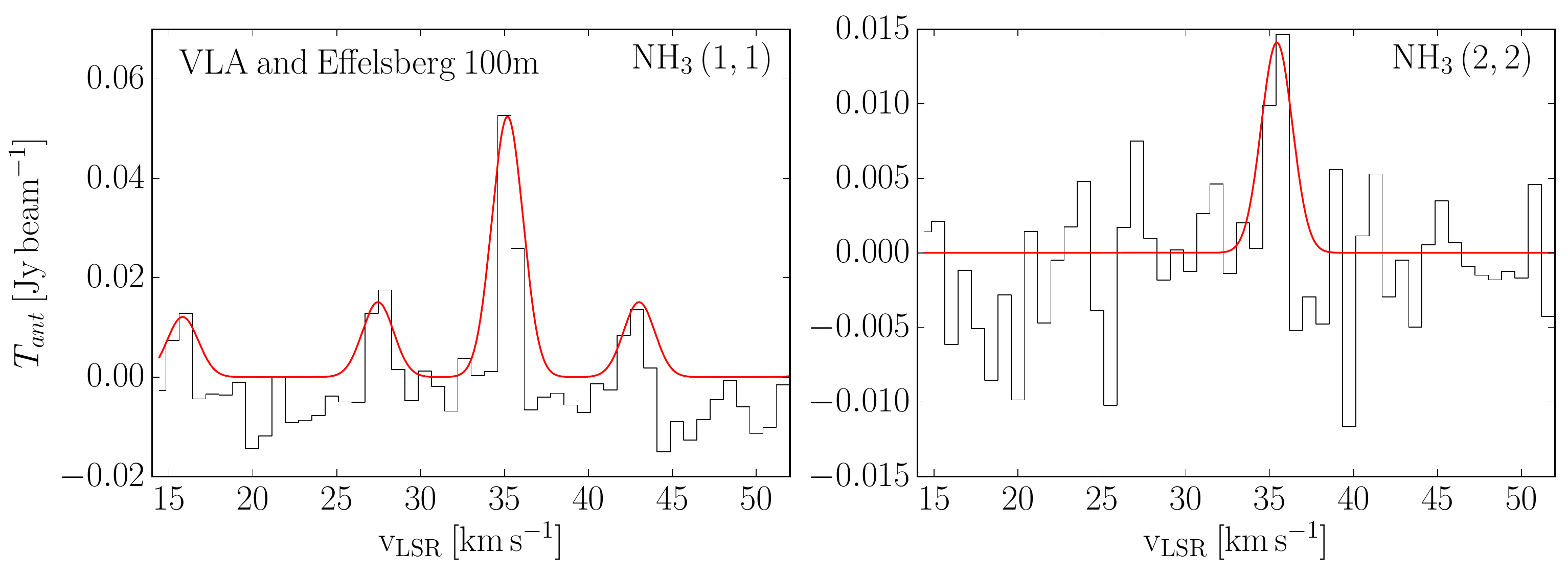}
       \caption{The NH$_3$ spectra of ISOSS18364. The left-hand side shows the NH$_3$(1,1) spectra and the right-hand side shows the NH$_3$(2,2) spectra. The top two panels show the Effelsberg 100m observations and the bottom panels show the combined VLA and Effelsberg 100m spectra at the position p1. The red line corresponds to the fit. As we do not observe the hyperfine structure of the NH$_3$(2,2) transition, we show a Gaussian fit to the main component. We do not detect the NH$_3$(2,2) line for the Effelsberg 100m observations within the uncertainties. The fit parameters and coordinates of the spectra are given in Table \ref{table_ammonia_spectra}.}
        \label{appendix_spectra_ISOSS18364}
\end{figure*}

\begin{figure*}
    \centering
       \includegraphics[width=17cm]{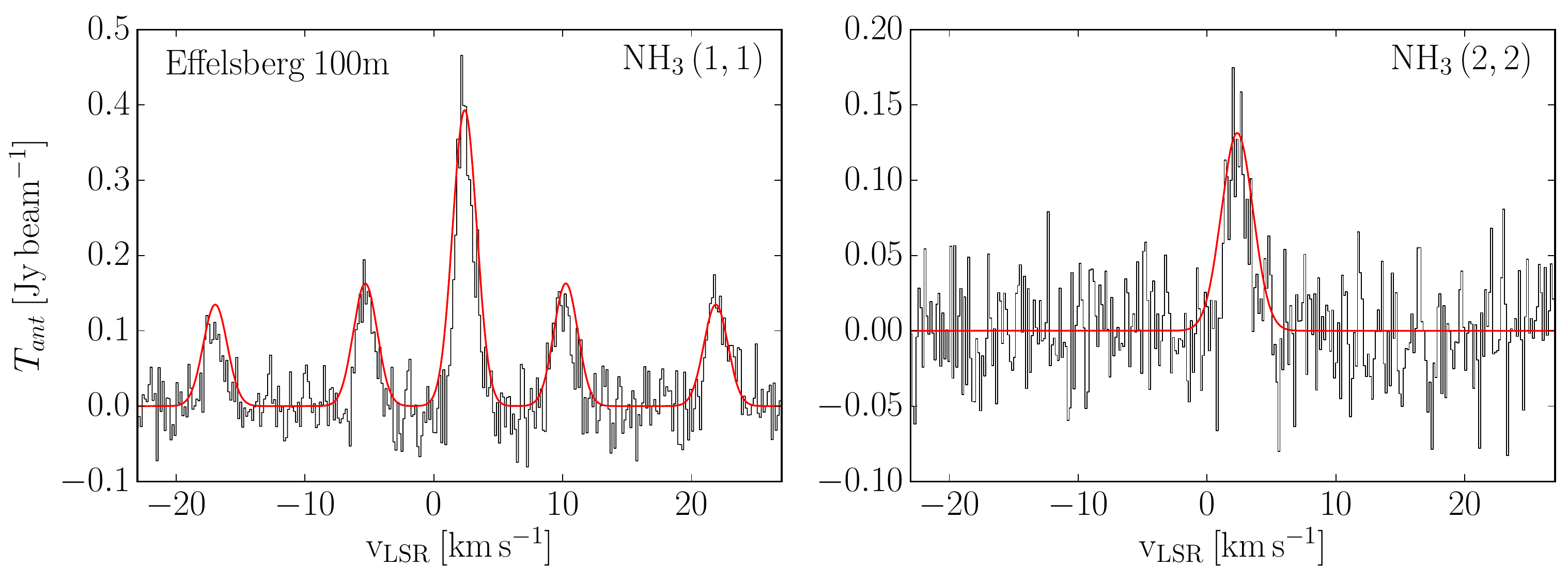}

       \includegraphics[width=17cm]{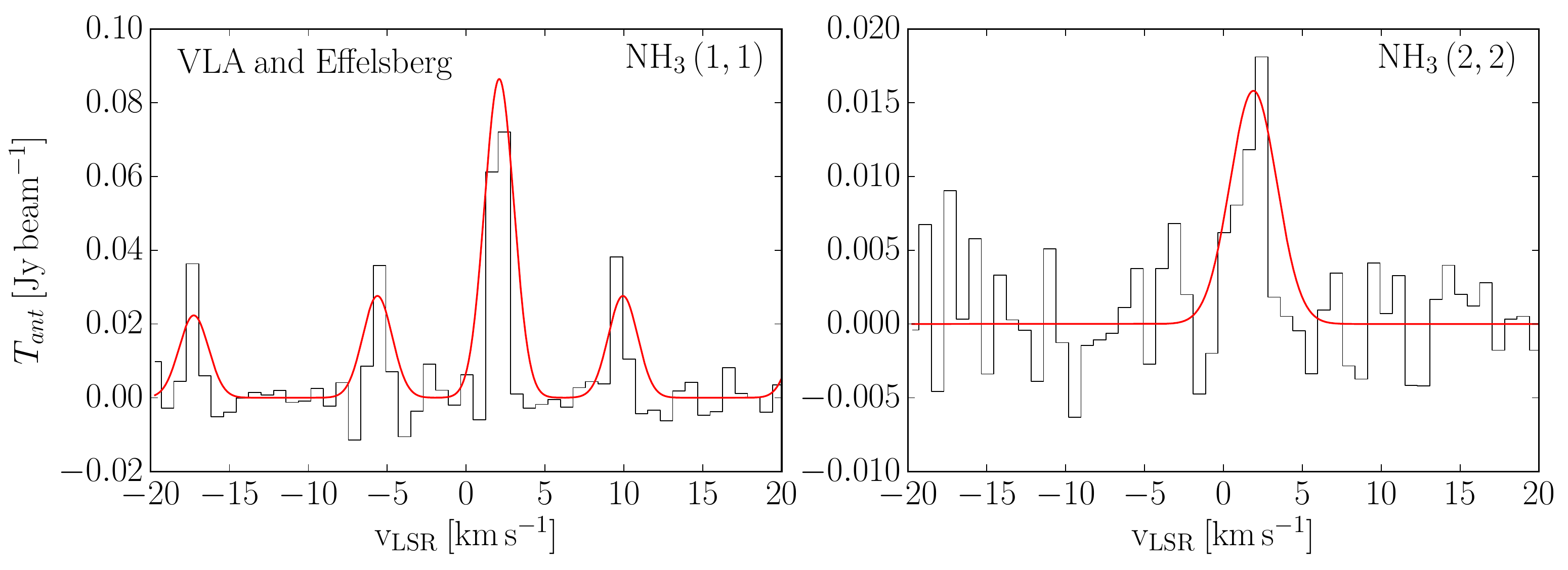}

       \includegraphics[width=17cm]{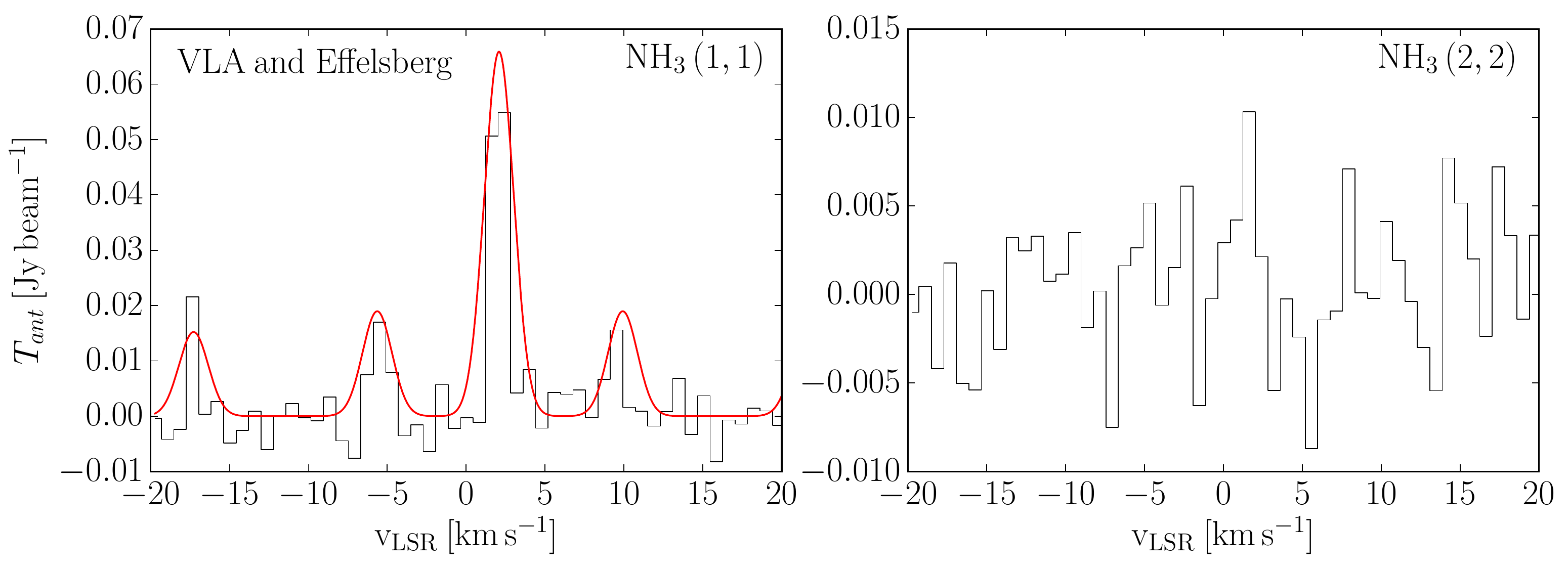}
       \caption{NH$_3$ spectra of ISOSS20153. The left-hand side shows the NH$_3$(1,1) spectra and the right-hand side shows the NH$_3$(2,2) spectra. The top two panels show the Effelsberg 100m observations, the central panels show the combined VLA and Effelsberg 100m spectra at the position p1, and the bottom panels show the combined VLA and Effelsberg 100m spectra at the position p2. The red line corresponds to the fit. Since we did not observe the hyperfine structure of the NH$_3$(2,2) transition, we show a Gaussian fit to the main component. We do not detect the NH$_3$(2,2) line at the position p2 within the uncertainties. The fit parameters and coordinates of the spectra are given in Table \ref{table_ammonia_spectra}.}
        \label{appendix_spectra_ISOSS20153}
\end{figure*}

\begin{figure*}
    \centering
       \includegraphics[width=17cm]{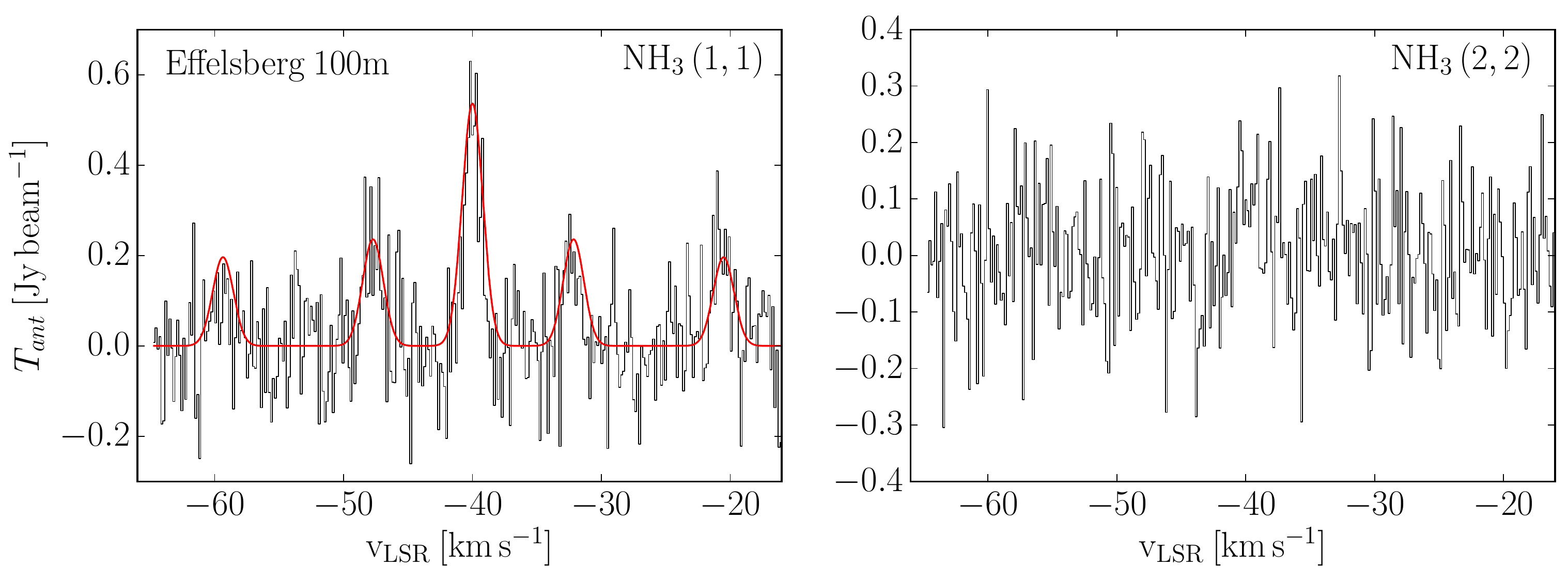}

       \includegraphics[width=17cm]{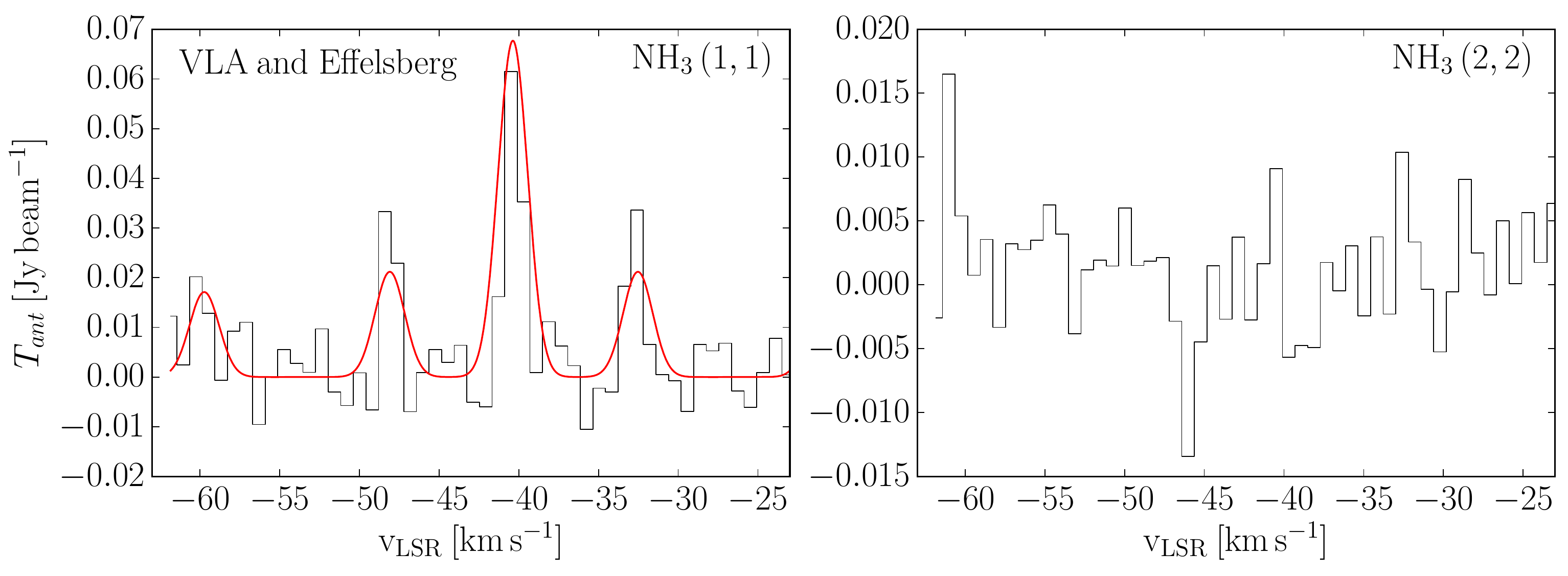}
       \caption{NH$_3$ spectra of ISOSS22478. The left-hand side shows the NH$_3$(1,1) spectra and the right-hand side shows the NH$_3$(2,2) spectra. The top two panels show the Effelsberg 100m observations and the bottom panels show the combined VLA and Effelsberg 100m spectra at the position p1. The red line corresponds to the fit. Within the uncertainties, we do not detect the NH$_3$(2,2) line. The fit parameters and coordinates of the spectra are given in Table \ref{table_ammonia_spectra}.}
        \label{appendix_spectra_ISOSS22478}
\end{figure*}

\begin{figure*}
    \centering
       \includegraphics[width=17cm]{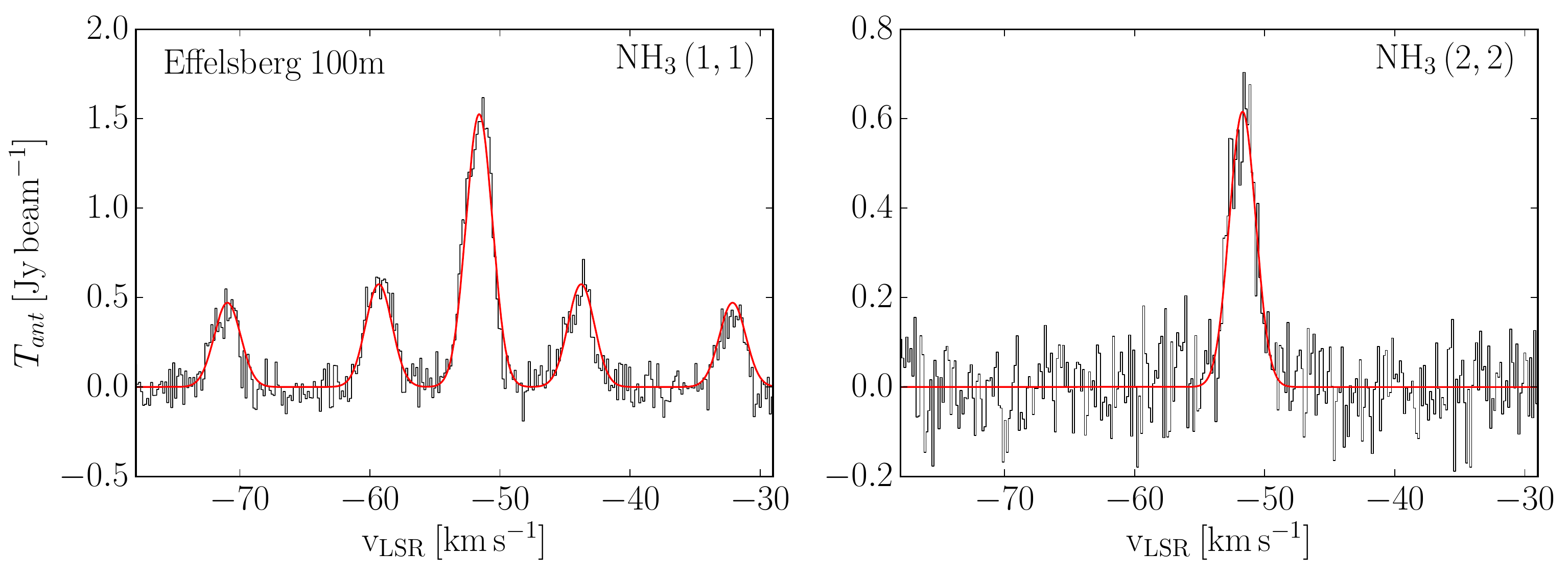}

       \includegraphics[width=17cm]{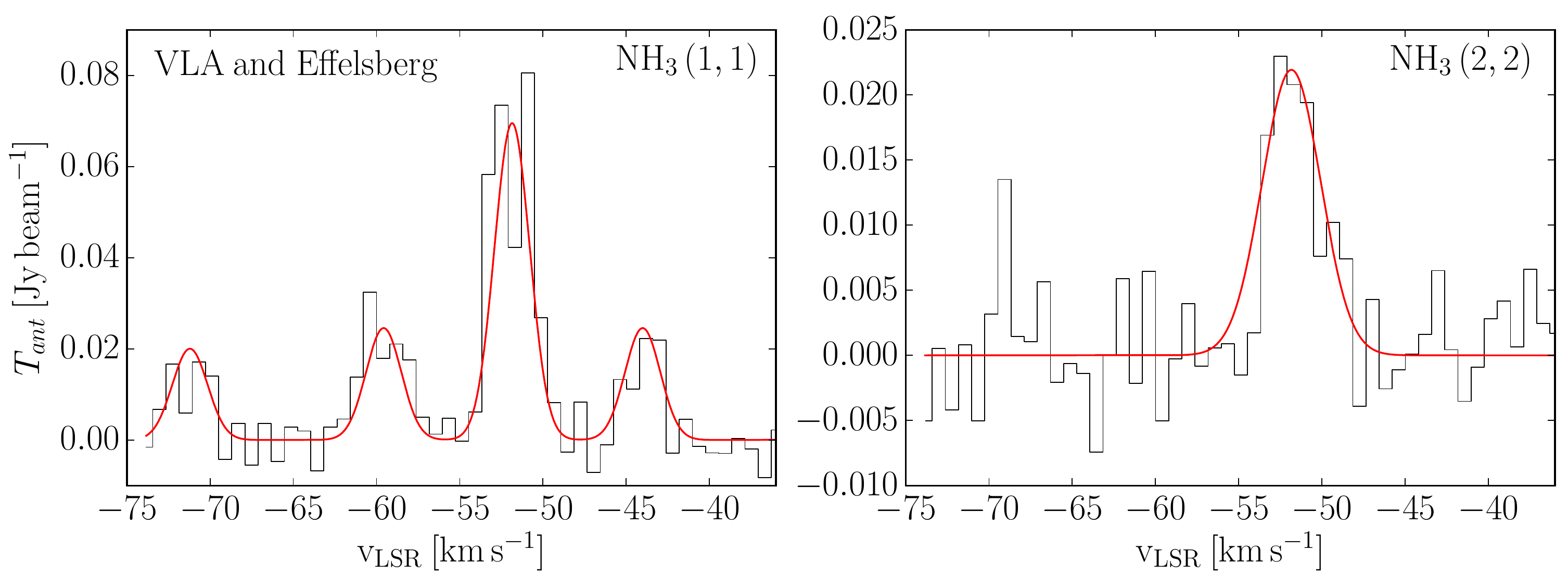}

       \includegraphics[width=17cm]{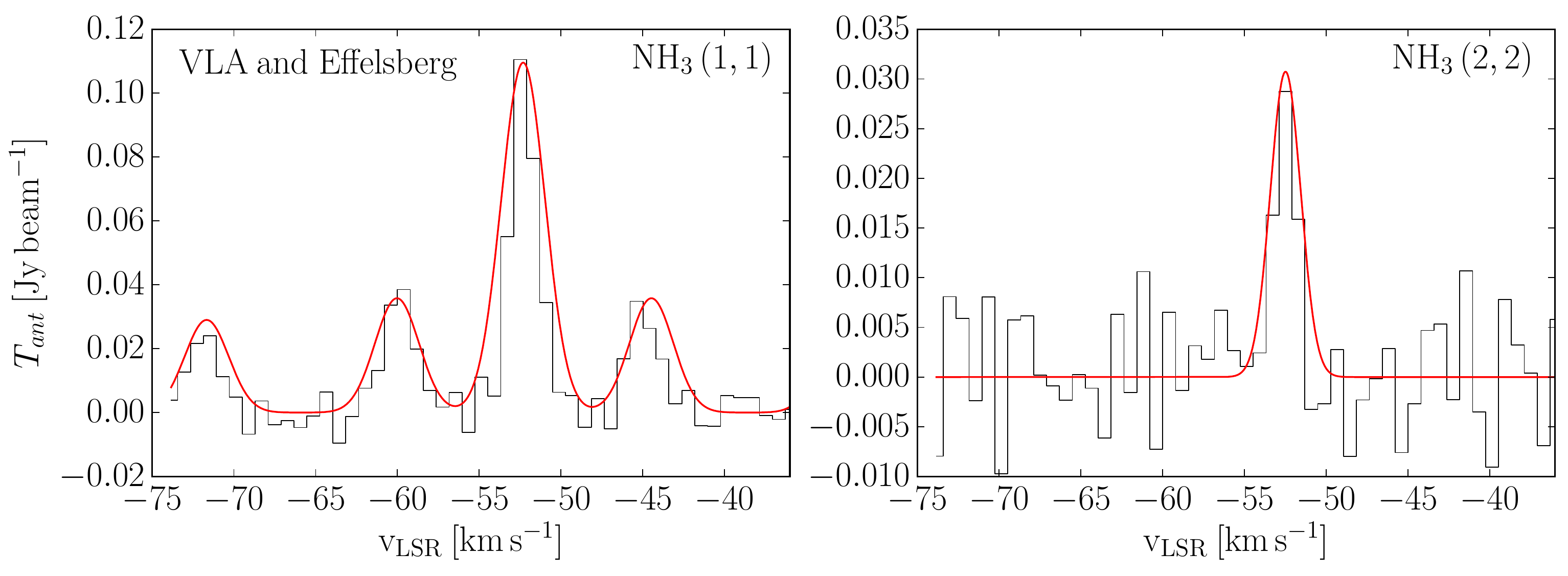}
       \caption{NH$_3$ spectra of ISOSS23053. The left-hand side shows the NH$_3$(1,1) spectra, and the right-hand side shows the NH$_3$(2,2) spectra. The top two panels show the Effelsberg 100m observations, the central panels show the combined VLA and Effelsberg 100m spectra at the position p1, and the bottom panels show the combined VLA and Effelsberg 100m spectra at the position p2. The red line corresponds to the fit. Since we do not observe the hyperfine structure of the NH$_3$(2,2) transition, we show a Gaussian fit to the main component. The fit parameters and coordinates of the spectra are given in Table \ref{table_ammonia_spectra}.}
        \label{appendix_spectra_ISOSS23053}
\end{figure*}

\end{appendix}

\end{document}